\documentclass[a4paper,11pt]{article}
\usepackage{pos}
\usepackage{times}
\usepackage[sort&compress, numbers]{natbib}

\let\oldbibliography\thebibliography
\renewcommand{\thebibliography}[1]{%
  \oldbibliography{#1}%
  \setlength{\itemsep}{6.67pt}%
}

\newcommand{\exclude}[1]{{}}
\renewcommand{\href}[1]{{\exclude{#1}}}

\newcommand{\pfrac}{$e^+/(e^+$$+$$e^-)$}

\newcommand{\gray}{$\gamma$-ray}

\newcommand{\fermilat}{{\it Fermi}--LAT}

\newcommand{\apj}{{\it ApJ}}

\newcommand{\apjs}{{\it ApJS}}
\newcommand{\aap}{{\it A\&A}}
\newcommand{\arnps}{{\it ARNPS}}

\newcommand{\prd}{{\it PRD}}

\newcommand{\adv}{{\it Adv.\ Spa.\ Res.}}

\newcommand{\jcap}{{\it JCAP}}
\newcommand{\mnras}{{\it MNRAS}}
\newcommand{\araa}{{\it ARA\&A}}

\newcommand{\pasa}{{\it PASA}}

\title{Direct measurements of cosmic rays \\ and their possible interpretations} 
\ShortTitle{Direct measurements of cosmic rays and their possible interpretations}

\author*{Igor V. Moskalenko}

\affiliation{Hansen Experimental Physics Laboratory
\& Kavli Institute for Particle Astrophysics and Cosmology, Stanford University, Stanford, CA 94305}

\emailAdd{imos@stanford.edu}

\abstract{
The last two decades have brought spectacular advances in astrophysics of cosmic rays (CRs) and space- and ground-based astronomy. Launches of missions that employ forefront detector technologies enabled measurements with large effective areas, wide fields of view, and precision that we recently could not even dream of. Meanwhile, interpretation of the individual slices of information about the internal working of the Milky Way provided by such experiments poses challenges to the traditional astrophysical models. New mysteries arise in the composition and spectra of CR species at low and high energies, in the energy range where we thought the main features were already understood fairly well. This accumulation of unsolved puzzles highlights the peculiarity of the current epoch and means that major breakthroughs are still ahead. In my talk, I review the current state of direct measurements of CRs and discuss their possible interpretations. Unfortunately, many important ideas and publications are not discussed here due to the space limitations.

}

\ConferenceLogo{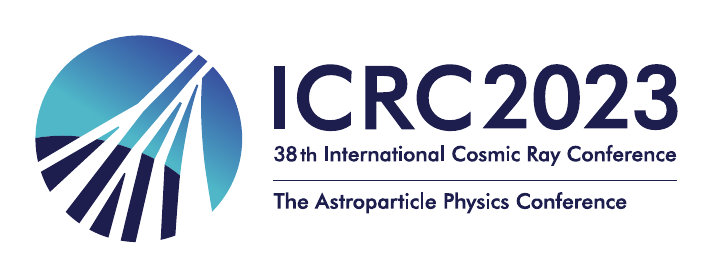}

\FullConference{%
38th International Cosmic Ray Conference (ICRC2023)\\
  26 July - 3 August, 2023\\
  Nagoya, Japan}

\begin{document}
\maketitle

\section{Introduction}

It is appropriate to start with the definition of the term ``direct measurements.'' The latter implies direct contact between CR particles and an instrument, which can only be done at the top of the atmosphere or in space. Apparently, the design, weight, and exposure of a scientific payload depend on the current technology and change over time, and so does the definition of the ``direct measurements.'' A plot of the all-particle CR spectrum, originally made by Simon Swordy in 2001 \cite{2001SSRv...99...85S}, and its well-known variations placed direct measurements below $\sim$1 TeV, reflecting the technology of the late 20th century, although there are several exceptions, such as, e.g., Sokol spacecraft \cite{1990YadF...51...99G, 1993RoIzF..57...76I} and JACEE experiment \cite{1998ApJ...502..278A}. Now, two decades later, we routinely make direct precise measurements approaching the energy of the knee at $\sim$3 PeV, thanks to the talents of experimentalists and recent technological advances. I enthusiastically predict that in about two decades, by 2050, direct measurements will advance the next three orders of magnitude in energy and reach 1-10 EeV range. Today, this can be considered a very challenging goal, but not impossible. 
Meanwhile, several breakthrough experiments have been proposed, e.g., AMS-100 \cite{2019NIMPA.94462561S}, HERD \cite{2023EPJWC.28001008P}, HERO \cite{2019AdSpR..64.2619K}, ALADInO \cite{2021ExA....51.1299B}, which significantly extend capabilities of the current instrumentation. For example, AMS-100 is designed to have a geometrical acceptance of $\sim$100 m$^2$ sr.

CR research is currently experiencing a golden age. Thanks to the outstanding progress in the energy coverage and precision of direct measurements, we are witnessing a constant stream of discoveries of new features in the spectra of CR species and anomalous behavior of elemental and isotopic ratios in the energy range from MV to 100s TV, and even more new experiments are preparing to launch (e.g., GAPS, HERD, TIGERISS, HERO, HELIX, COSI). A natural consequence is an \emph{infinite} number of publications with interpretations of new observed features. Unfortunately, this means that many important ideas and publications are not mentioned here due to space limitations.
 
For most of the 20th century, the two popular models describing the observed spectra and composition of CRs were the leaky-box and the Galactic diffusion model with halo \cite{1990acr..book.....B}. The leaky-box model is the simplest and some people still use it. This model considers the Galaxy as a volume uniformly filled with gas, sources, and CRs with a small leakage -- hence the name ``leaky-box.'' Tuned to local measurements, it can correctly reproduce the fluxes of stable nuclei at a single point in the Galaxy. The diffusion model is a realistic model in which gas and sources are distributed in the Galactic disk. CRs fill a large volume around the disk called the halo, and can escape into the intergalactic space through its outer ``boundaries.'' The spatial distributions of CR species, gas, background radiation, and magnetic field are non-uniform and generally consistent with observations of diffuse Galactic thermal and non-thermal emissions. This model 
is widely used today.

At the turn of the 21st century, the general success of the diffusion model led Vitaly Ginzburg to the conclusion \cite{1999PhyU...42..353G}: ``In respect of CR with $E_{\rm CR}$$<$$10^{15}$$-$$10^{16}$ eV, there generally remain some vague points, but in the whole the picture is \emph{clear} enough...'' This is reminiscent of the popular view at the turn of the 20th century (ca.\ $\sim$1900), formulated by Lord Kelvin (William Thomson): ``There is nothing new to be discovered in physics now. All that remains is more and more precise measurement...'' Interestingly, increasingly precise measurements is exactly what led us to the current situation! Today we have such excellent data that the whole picture becomes rather \emph{unclear}. This creates exciting opportunities for theorists and experimentalists and promises new breakthroughs!


A word of caution though. A large amount of new material and attempts to comprehend it often lead to the development of overly complicated models, which add many parameters in an attempt to achieve consistency with the data. Such models are featuring two halos with different diffusion coefficients, anisotropic diffusion, local sources with different injection spectra, sources in spiral arms, slow diffusion zones, etc., sometimes combined within a single model. Although this allows models to reproduce some of the observed features, it does not necessary lead to a better understanding of the underlying processes. Here we must adhere to time-tested wisdom, such as that formulated by William of Ockham (known as Occam's razor): \emph{Numquam ponenda est pluralitas sine necessitate} (``Plurality must never be posited without necessity'') or follow the advise attributed to Albert Einstein: ``Everything should be as simple as it can be but not simpler.''

\section{Low-energy features}

Low-energy measurements provide unique information about isotopic spectra and composition of CRs. However, such measurements are subject to heliospheric (or solar) modulation, which was difficult to handle due to the lack of measurements outside the heliosphere. This problem was largely resolved with Voyager 1, 2 entering the very local interstellar medium (ISM), V1 in 2012 and V2 in 2018, and beaming the elemental spectra of CR species from ISM space \cite{2016ApJ...831...18C}. This unimaginable breakthrough provides a solid ground for studies of low-energy particles.  

Low-energy features, often called excesses, are observed in the spectra of some elements when compared to the spectra of their neighbors, there are also mismatches between the data taken by different instruments operating in different energy ranges, even after their correction for solar modulation, and/or significant deviations from model predictions. Let us start with iron and its radioactive isotope, $^{60}$Fe ($\beta^-$ decay, half-life of 2.6 Myr \cite{2009PhRvL.103g2502R}), which may provide some clues to the origin of low-energy features in the spectra of other species.


\subsection{$^{60}$Fe as a tracer of supernova activity in the solar neighborhood}

Evidences of the past SN activity in the local ISM are abundant \cite{2015ApJ...800...71F, 2016Natur.532...69W, 2016Natur.532...73B}. There is no general agreement on the exact number of SN events and their exact timing, but it seems quite clear that several events may have occurred at distances of up to 100 pc in the last $\sim$10 Myr. The most recent SN events in the solar vicinity occured 1.5--3.2 Myr and 6.5--8.7 Myr ago \citep{2015ApJ...800...71F, 2016Natur.532...69W}. The measured signal spread of $\sim$1.5 Myr implies a series of SN explosions.
Besides, the Local Bubble is a low-density region about $\sim$200 pc around the sun filled with hot H\,\textsc{ii} gas, itself formed by a series of SN explosions \cite{1999A&A...346..785S, 2011ARA&A..49..237F}. Studies suggest 14--20 SNe within a moving group, the surviving members of which are now in the Scorpius-Centaurus stellar association \cite{2011ARA&A..49..237F, 2016Natur.532...73B}.


An excess of radioactive $^{60}$Fe found in deep ocean core samples of FeMn crust \citep{1999PhRvL..83...18K, 2004PhRvL..93q1103K, 2016Natur.532...69W}, in the Pacific Ocean sediments \citep{2016PNAS..113.9232L}, in lunar regolith samples \citep{2009LPI....40.1129C, 2012LPI....43.1279F, 2014LPI....45.1778F}, and in the Antarctic snow \citep{2019PhRvL.123g2701K} indicates that it may have been deposited by SN explosions in the solar neighborhood. 
Fifteen $^{60}$Fe events and only one $^{61}$Co event were observed by ACE-CRIS \citep{2016Sci...352..677B} in CRs, while there were about equal numbers of $^{58}$Fe and $^{59}$Co events. This implies that the $^{60}$Fe events are real and not the spillover of a more abundant neighboring isotope. Meanwhile, only an upper limit was established for $^{59}$Ni ($\tau_{1/2}$$\sim$76 kyr) \citep{1999ApJ...523L..61W}, suggesting $\gtrsim$100 kyr time delay between the ejecta and the next SN. 

The low-energy feature in the \emph{iron spectrum} \cite{2021ApJ...913....5B}, perhaps also associated with SN activity in the solar neighborhood, was revealed for the first time when comparing data from Voyager 1 \cite{2016ApJ...831...18C} and ACE-CRIS with AMS-02 \cite{2021PhRvL.126d1104A}. It is most clearly visible as a bump in the Fe/He, Fe/O, and Fe/Si ratios at 1--3 GV, while a similar feature in the He/O and Si/O ratios is absent. The large fragmentation cross section and fast ionization losses of iron hint at a local origin of the excess.
The calculations \cite{2021ApJ...913....5B} use the Monte Carlo code HelMod \cite{2019AdSpR..64.2459B} based on the Parker equation and developed to describe the CR transport through the heliosphere from interstellar space to the Earth.
 
Interestingly, the Ni/Fe ratio reported by CALET \cite{2022PhRvL.128m1103A} (see also a highlight CALET talk by Shoji Torii \cite{Torii:20231Q}) is constant between 10 and 200 GeV/n, indicating the same origin of elements of the iron group. Precise measurements of the sub-Fe/Fe\,=\,(Sc+Ti+V)/Fe ratio can shed light on the origin of the iron group and provide further details about CR sources in our local neighborhood.

\subsection{Aluminum excess}

The excess in the Al spectrum in the narrow rigidity range of 3--10 GV ($\sim$0.8--4 GeV/n) becomes clearly visible if we compare the Al/Si ratio measured by AMS-02 \cite{2021PhRvL.127b1101A} with the model predictions \cite{2022ApJ...933..147B}, while a similar feature in the Na/Si ratio is absent. There are four possible physical reasons for the discrepancy between data and model calculations \cite{2022ApJ...933..147B}: (i) incorrect spectrum of $^{28}$Si, the main progenitor of secondary $^{26,27}$Al, (ii) errors in the total Al inelastic cross sections, (iii) errors in the production cross sections of $^{26,27}$Al isotopes, and (iv) additional local component of primary Al. 

Reason (i) can be rejected because all model calculation are based on available data. The Si spectrum is tuned to data from Voyager 1, ACE-CRIS, and AMS-02 \cite{2020ApJS..250...27B}. Importantly, AMS-02 data is available above 2 GV, the same rigidity range for all CR species. Contributions of other CR species are very minor and cannot be the cause of the observed excess. 
(ii) Significant errors in the total inelastic cross section of Al can be excluded as the primary cause of the excess, taking into account the accelerator data. The total inelastic cross section of $^{27}$Al is measured below $\sim$1 GeV/n ($<$3 GV) \cite{BAUHOFF1986429} and in the rigidity range 10--19 GV in the inverse kinematics \cite{Bobchenko:1979hp}. The parameterizations of the total inelastic cross sections used in the model calculations are tuned to the available data. One can also notice the absence of similar excesses in the spectra of neighboring nuclei, such as Ne, Na, Mg, Si \cite{2020ApJS..250...27B, 2022ApJ...933..147B}.
(iii) Almost 100\% of secondary $^{27}$Al is produced through fragmentation of $^{28}$Si with minor contributions from $^{29}$Si, $^{32}$S, and $^{56}$Fe \citep{2013ICRC...33..803M}. Unfortunately, the isotopic production cross section ($p+^{28}$Si$\to ^{27}$Al) is the major source of uncertainty. Only a couple of data points are available for this reaction, at 1.4 GV and 2.3 GV in the inverse kinematics, which constrain $^{27}$Al production in the lower range of rigidities where the excess is observed. Fortunately, measurements of $^{28}$Si fragmentation $^{28}$Si$+p$$\to$Al performed by a Siegen group \citep{1991ICRC....2..280H} indicate that the cross section remains flat between 1 and 14.5 GeV/n corresponding to the rigidity range from 3.5--31 GV ($^{27}$Al), which covers the excess range and extends to significantly higher rigidities. (iv) The observed low-energy excess in the Al spectrum is most likely due to the stable isotope $^{27}$Al. However, because it is stable, its yield and Galactic distribution are difficult to measure. Meanwhile, an abundant literature exists on observations of the distribution of the diffuse \gray{} 1.809 MeV line emission from the decay of the radioactive $^{26}$Al isotope and its origin. 

Observations of 
the diffuse Galactic 1.809 MeV emission line by COMPTEL \citep{1999A&A...345..813K} and INTEGRAL \citep{2015ApJ...801..142B} have shown that $^{26}$Al nucleosynthesis is ongoing in the present Galaxy. Potential sources include AGB stars, novae, core-collapse SNe, and Wolf-Rayet stellar winds \citep{1996PhR...267....1P, 2006Natur.439...45D}. There are also reports of the discovery of a close SNR (G275.5+18.4) with an  angular diameter of $24^\circ$ in the constellation Antlia Pneumatica \cite{2002ApJ...576L..41M}. The SNR distance is estimated as 60--340 pc, but is most likely $\sim$200 pc \cite{2022ApJ...940...63R}. A marginally significant feature is detected in the 1.8 MeV \gray{} emission line within the Antlia SNR.

\subsection{Lithium excess}

The standard propagation model \cite{2020ApJ...889..167B} tuned to the B/C ratio data by Voyager 1  \citep{2016ApJ...831...18C}, ACE-CRIS \citep{2013ApJ...770..117L}, and AMS-2 \cite{2018PhRvL.120b1101A}, demonstrates good agreement with measurements of CR species in a wide energy range, implying that Li spectrum should also be well reproduced by the same model. However, a comparison of the model calculations of secondary Li with data exhibits a significant excess over the model predictions above a few GV \cite{2020ApJ...889..167B}. This may be an indication of errors in the production cross sections or an unexpected primary Li component.

From a compilation of the majority of Li production cross sections \cite{2018PhRvC..98c4611G}, one can see that the main production channels are the fragmentations of $^{12}$C and $^{16}$O 
measured in several different experiments. Although they are not measured perfectly, each contributes 12\%-14\%, and thus a 20\% error in one of them would correspond to only 2\%-3\% of total Li production. Other production channels contribute at a level of 1\%-2\% or less. It is not impossible, but rather unlikely that the cross section errors are all biased in the same direction, resulting in the observed 20\% excess.

The obvious ``solution'' is to renormalize the LiBeB production cross sections to match the CR data and/or reduce the diffusion coefficient to boost Li production, while ensuring the calculated B/C ratio is still consistent with the experimental data \cite{2020A&A...639A.131W, 2021JCAP...07..010D}. The scientific terms for this approach are scale factors and nuisance parameters. While these may seem cool and produce the desired results, hastily removing the excess risks throwing the baby out with the bathwater.

Another cross section hypothesis suggests that the contribution of Fe fragmentation is calculated incorrectly, and with new updated cross sections, the Li data can be reproduced well enough \cite{2022A&A...668A...7M}. Meanwhile, a back-of-the-envelope estimate shows that the contribution of Fe to Li production is less than 5\% at 22 GV (at maximum excess). 

Indeed, the \emph{local ISM fluxes} of main progenitors taken at $E_{\rm kin}$=10.44 GeV/n ($\approx$22 GV) are: $F_{\rm Fe}$$\approx$$4.476\times10^{-3}$, $F_{\rm Si}$$\approx$$6.723\times10^{-3}$, $F_{\rm Mg}$$\approx$$7.904\times10^{-3}$, $F_{\rm Ne}$$\approx$$6.456\times10^{-3}$, $F_{\rm O}$$\approx$$4.131\times10^{-2}$, $F_{\rm N}$$\approx$$9.089\times10^{-3}$, $F_{\rm C}$$\approx$$3.979\times10^{-2}$, $F_{\rm B}$$\approx$$7.709\times10^{-3}$ in units of m$^{-2}$ s$^{-1}$ sr$^{-1}$ [GeV/n]$^{-1}$, where the fluxes of B--Si are taken from \cite{2020ApJS..250...27B}, and Fe -- from \cite{2021ApJ...913....5B}. The heliospheric transport in \cite{2020ApJS..250...27B, 2021ApJ...913....5B} is calculated using the HelMod code, but the solar modulation at $\approx$22 GV is weak anyway.
At 10 GeV/n, the cross sections of the reactions $A+p$$\to$Li\,$+X$, where $A=$ B, C, N, O, Ne, Mg, Si, Fe, and Li=$^6$He+$^6$Li+$^7$Li, are about the same in the range of 23--32 mb, see Figs.\ 2, 3 in \cite{2022A&A...668A...7M}. Assuming they are \emph{all the same}, we can simply use fluxes $F_A$ shown above. This gives for the iron contribution $F_{\rm Fe}/\Sigma_A F_A$$\approx$$0.036$ (3.6\%), where  $\Sigma_A F_A$$\approx$$0.123$. Even if the Fe$\to$Li cross section
is 50\% larger (unlikely), the Fe contribution increases to $\approx$$0.053$ (5.3\%).  These numbers are calculated without taking into account the contribution of all other species, so the Fe contribution is even smaller in reality. Note that the effect of possible cross section errors in reactions with He target is minor \cite{2022A&A...668A...7M}.
Although the contribution of Fe increases with rigidity due to its hard spectrum, this estimate shows that incorrect cross sections cannot be the main cause of the excess.

An exciting possibility is that the primary Li may come from nova explosions \cite{2020ApJ...889..167B}. Indeed, the $\alpha$-capture reaction of $^7$Be production $^3$He$(\alpha,\gamma)^7$Be in stars was proposed a while ago \citep{1955ApJ...121..144C, 1971ApJ...164..111C}. A subsequent decay of $^7$Be (half-life of 53.22 days) yields $^7$Li isotope. $^7$Be should be transported into cooler layers where it can decay to $^7$Li, the so-called Cameron-Fowler mechanism in AGB stars. The production of $^7$Li in the same reactions in novae was discussed in \cite{1975A&A....42...55A,1978ApJ...222..600S, 1996ApJ...465L..27H}. 
Observation of blue-shifted absorption lines of partly ionized $^7$Be in the spectrum of a classical nova V339 Del \citep{2015Natur.518..381T} about 40-50 days after the explosion is the first  evidence that the mechanism proposed in 1970s is working indeed \citep{2015Natur.518..307H}. Consequent observations of several other novae (V1369 Cen, V5668 Sgr, V2944 Oph, V407 Lupi, V838 Her) also reveal the presence of $^7$Be lines in their spectra. Meanwhile, there could be other sources of $^{6,7}$Li. 
Low-energy measurements by PAMELA \cite{2018ApJ...862..141M} show that $^7$Li/$^6$Li$\approx$1.1 below 1 GeV/n. The \emph{latest} preliminary AMS-02 analysis \cite{Wei:20231Q} indicates $^7$Li/$^6$Li$\approx$1 below 10 GeV/n, but this still may change when the analysis is completed.
Precise measurements of the isotopic ratio can shed light on the origin of Li in CRs.

\section{Inventory of Galactic cosmic ray sources}

The main sources of Galactic CRs remain SNe and SNR with their total kinetic energy of the ejecta in the range of $10^{51}$ erg. The Green catalog currently lists 294 SNR. Meanwhile, the isotopic and spectral anomalies observed recently force us to look at other sources, especially local, which can also contribute to the observed CRs. These are primarily Wolf-Rayet stars (currently 354 are known) and O-stars (20,000 observed), which over their fairly short lifetimes provide, respectively, $10^{51}$ erg and $10^{50}$ erg in high-velocity winds reaching $(2-4)\times10^3$ km/s, pulsars ($\sim$1,500 observed) with their total rotating power reaching $4\times10^{49}$ erg (Crab), and novae providing $10^{45}$ erg (estimated frequency 30--50/year, $\approx$350 observed). For comparison, countless stellar flares can provide up to $10^{36}$ erg each, and can also add to low-energy CRs.   

\section{Protons/Helium ratio}


The monotonic decrease of the H/He ratio was first noticed in the PAMELA data when plotted vs.\ rigidity \cite{2011Sci...332...69A} and has been confirmed by other experiments covering the impressive rigidity range from $\sim$100 MV to 50 TV, such as Voyager 1 \citep{2016ApJ...831...18C}, AMS-02 \cite{2021PhR...894....1A} (Fig. 79), and CALET \cite{Adriani:2023E5}, while DAMPE \cite{Coppin:2023Px} presents only their latest proton and He spectra, but not their ratio.
Meanwhile, we note that a flatter spectrum of He vs.\ H was already spotted by many earlier experiments, such as Sokol \cite{1990YadF...51...99G, 1990ICRC....3...77I}, JACEE \cite{1998ApJ...502..278A}, ATIC \cite{2009BRASP..73..564P}, and CREAM \cite{2010ApJ...714L..89A}.
However, theory told us that the spectral indices in rigidity should not depend on the specific properties of the primary particles. Thus, many researchers then have attributed this difference to systematic effects, and I myself am no exception. 

Apparently, not only He behaves differently from protons. Accurate measurements by AMS-02 \cite{2021PhR...894....1A} show that other, mainly primary species, such as C, O, Si, have spectral indices above $\sim$60 GV very similar to He. This raises a question, whether the spectrum of primary CR species depends on the $A/Z$ ratio, which is equal to 1 for protons, and 2 for most abundant isotopes of He, C, O, Si.

\subsection{Hypothesis of the spatial distribution of elements}

The main idea is that SN explodes into the pre-SN wind, which is composed of lighter elements when the star is young, but becomes increasingly enriched in heavier elements in its later stages \cite{2011ApJ...729L..13O, 2016PhRvD..93h3001O}. The young SN shell then accelerates heavier elements when it is young, and lighter elements when it fates. This would make the spectra of heavier species flatter as they are accelerated by a stronger shock, while the spectra of lighter elements are produced at later stages. SN can also explode into the medium enriched with heavier elements from previous SN explosions. 

An argument against this hypothesis is that the spectra of He, C, and O have the same index, while the spectra of Ne, Mg, and most importantly Si are somewhat steeper \cite{2021PhR...894....1A}. This implies that the spatial distribution of C and O in the pre-SN wind should match the distribution of He, while heavier Ne, Mg, and Si should be accelerated by a weaker shock at a later time.

\subsection{Hypothesis of two components in the H spectrum}

This is an empirical hypothesis \cite{2021PhR...894....1A}, which suggests that the observed CR proton spectrum is a combination of spectra, which come from two distinctly different types of proton sources. One of them is a regular source that accelerates all particles and injects them into the ISM with spectra similar to He. Another source is enriched with hydrogen (or depleted in heavier species) and injects protons with a steeper spectrum (by $\approx$0.3 in index).

Earlier, a similar idea \cite{2015ApJ...803L..15T, 2019PhRvD.100f3020Y} was proposed to reproduce the observed positron excess. The harder spectrum (younger) sources should be surrounded by gas producing more secondary species including positrons, and this could explain their flatter spectrum.

An argument against this hypothesis is that it requires sources that are unique for protons. It is not clear what kind of sources these are, and what makes them so unique.  

\subsection{Hypothesis of different acceleration efficiency}

The ideas proposed in \cite{2019ApJ...872..108H} and \cite{2017PhRvL.119q1101C} are somewhat different, but the authors came to similar conclusions: most particles in the shock are protons ($A/Z$$=$$1$), which generate Alfv\'en waves and became frozen into the generated turbulence. The nuclei with $A/Z$$>$$1$ or $A/Q$$>$$1$ (Q is the charge of a partly ionized atom) are not in synch with Alfv\'en waves generated by protons, and are more efficiently injected into the shock and then accelerated. 

The hypothesis predicts that the injection efficiency of heavier species increases relatively to protons with increase of the shock Mach number and $A/Q$ value. The same applies to all species $A/Q$$>$$1$, but the efficiency should come to saturation for sufficiently high $A/Q$ values.

\section{Silicon \& fluorine puzzles}\label{Si-F}


An increase in accuracy of CR data reveals unexpected puzzles in the O-Si groups, implying that CR acceleration and transport processes are still far from being fully understood.
Let us look at the Si/O ratio in the local ISM \cite{2021ApJ...913....5B}, i.e.\ take their local interstellar spectra \cite{2020ApJS..250...27B}, which eliminates the solar modulation. The most abundant isotopes, $^{16}$O and $^{28}$Si, are primaries with $A/Z$$=$$2$. The larger fragmentation cross section and faster ionization energy losses of Si nuclei result in the rise of the Si/O ratio with rigidity below $\sim$10 GV.
In the absence of energy losses in the middle range ($\sim$10--300 GV), the Si/O ratio is const, while at higher rigidities, the ratio decreases for no apparent reason. Note that the He/O ratio is flat above 60 GV.


It is also interesting to examine the differences in two secondary/primary ratios, B/O and F/Si. If we look at the rigidities $>$10 GV, where the solar modulation is small, the ratios diverge in the entire range up to 1 TV with the F/Si ratio being flatter by $\approx$0.052 in the index (Fig.\ 3 in \cite{2021PhRvL.126h1102A}), albeit with large error bars in the F/Si ratio. This usually implies a difference in the interstellar propagation, with the indices of the effective diffusion coefficients probed by these ratios differing by that same number. Then, a weak monotonic increase can be expected for the F/B ratio. However, the increase is observed only up to 100--200 GV, while in the range $\sim$100 GV--1 TV the ratio becomes constant (Fig.\ 1 in \cite{2021PhRvL.126h1102A}), again with large error bars. The break at $\approx$200 GV, if confirmed, perhaps has the same origin as the break observed in the spectra of all CR species and discussed in the next section. Striking is that the increase in the F/B ratio is observed in the range 10--200 GV, where Si/O$\approx$const. The latter defies the hypothesis of the difference in propagation, as in this case, the Si/O ratio cannot be constant, assuming that the injection spectra of O and Si are the same. 



Let us look at the fluorine puzzle from a different perspective. A comparison of standard propagation calculations tuned to the B/C or B/O ratio with the measured F/Si ratio shows a deficit in secondary fluorine, which increases with the rigidity up to the break at 200 GV \cite{2022ApJ...925..108B}. However, it becomes consistent with the AMS-02 data \cite{2021PhRvL.126h1102A} and thus with the B/O ratio above 200 GV, albeit with large error bars. This is a serious issue, which cannot be cured by simple renormalization of the cross sections -- the latter are flat above $\sim$1--2 GeV/n. For example, if we assume that the fluorine production cross sections are off by $\approx$10\% and renormalize the calculated fluorine spectrum down by the same factor, we obtain excesses from $\approx$3--10 GV and at $\sim$100 GV.

The described rigidity-dependent discrepancies imply different origins of the Si group and the CNO group, or an as yet unclear difference in their propagation, or perhaps a non-negligible primary F component. The F anomaly is also confirmed in other studies, e.g., \cite{2023PhRvD.107f3020Z}. Precise measurements of the $_{15}$P/$_{16}$S and sub-Fe/Fe ratios should be able to clarify the issue or add more puzzles.

\section{200 GV \& 10 TV breaks or the TV bump}

The $\sim$200 GV breaks in the spectra of protons and He are clearly visible in the data collected by  ATIC-2 \cite{2009BRASP..73..564P}, CREAM \cite{2010ApJ...714L..89A}, and even earlier experiments, but initially looked like a calibration problem between lower and higher energy ($\lessgtr$200 GeV) experiments. The break becomes widely accepted after the PAMELA publication \cite{2011Sci...332...69A}, and a confirmation by Fermi-LAT \cite{2014PhRvL.112o1103A}, which used observations of the CR-induced $\gamma$-ray emission from the Earth's limb. The latter actually confirmed the flatter proton spectrum above $\sim$200 GeV, in agreement with PAMELA measurements.



The break is most clearly visible in the latest data by AMS-02 \cite{2021PhR...894....1A}, the instrument that is best suited for this energy range. A comparison of the fits made in the range from 30-50 GV to 200 GV shows an interesting picture, where Fe has the hardest spectrum followed by He, O, C, and then Si, S, Ne, Mg. The steeper spectra are observed in H, Al, N, Na, F, B, Be, and the steepest is Li (partly tertiary). The fluorine spectrum is flatter than the spectrum of boron, as already discussed above, and may indicate a different origin or presence of the primary component.

Yet another break, at 10 TV, was also observed by the ATIC \cite{2009BRASP..73..564P} and CREAM \cite{2017ApJ...839....5Y} teams, but the former made no claim, while the latter stated that more data is needed. The first decisive evidence was provided by the NUCLEON team \cite{2018JETPL.108....5A}, which observed the break in the spectra of protons, He, and light elements at the same rigidity. This break is now confirmed by several instruments: DAMPE \cite{2019SciA....5.3793A}, CALET \cite{2022PhRvL.129j1102A}, and ISS-CREAM \cite{2022ApJ...940..107C}. A striking increase in anisotropy in the same energy range from $\sim$0.2--100 TeV \cite{2019ApJ...871...96A} indicates that these two breaks form a single bump structure from $\sim$0.2--100 TV rather than being two independent features; note that for protons dominating in CRs, their kinetic energy is approximately equal to the rigidity at high energies. 

\subsection{Anisotropy map}

The anisotropy map of the entire sky at 10 TeV, which corresponds to the bump maximum, was produced using the combined data of HAWC and IceCube experiments  \cite{2019ApJ...871...96A}. It includes the large-scale and small-scale anisotropy features, and provides information that is crucial for understanding the bump origin. The large-scale map features a very sharp jump in the relative CR intensity across the magnetic equator -- a hint at the proximity of the source. The dominant spot in the residual small-scale anisotropy map (Region A, Figs. 5, 11 \cite{2019ApJ...871...96A}) points to the direction to the source, which coincides with the Galactic \emph{anti-center}, the direction of the local B-field, and is about $45^\circ$ off the ``tail'' of the heliosphere. This is in a remarkable contradiction with the conventional understanding that the phase of dipole anisotropy should point to the direction of the Galactic center, where the majority of CR sources are located and the CR number density is the highest.


\section{Models of the TeV bump}

On March 3, 2011, PAMELA reported about a break in the spectra of H and He at about the same rigidity 230--240 GV \cite{2011Sci...332...69A}. No one knew yet that there was another break at higher rigidity. Therefore, the early papers proposed interpretation of this first break. The first paper \cite{2012ApJ...752...68V}, submitted on August 4, 2011, proposed four different scenarios of the origin of the break: interstellar propagation, the injection spectrum (or spectra from two distinctly different source populations), and a local source at low or high energies. 
The analysis showed that the propagation scenario in which the break is associated with a change in the rigidity dependence of the diffusion coefficient is preferable.
These scenarios and their variations
are still discussed in the literature. 
Physical interpretation for the propagation scenario \cite{2012PhRvL.109f1101B} was proposed 10 months later (submitted on May 30, 2012): ``...the diffusive propagation is no longer determined by the self-generated turbulence, but rather by the cascading of externally generated turbulence (for instance due to supernova bubbles) from large spatial scales to smaller scales where CRs can resonate."


The propagation scenario naturally explains the same break rigidity for all species and reproduces the observed difference between the spectra of primary and secondary species in subsequent AMS-02 publications. Essentially, the values of the spectral breaks in the spectra of primary (C, O) and secondary (B) species are connected as $\Delta \delta_{\rm sec}$$\approx$$2\Delta \delta_{\rm pri}$, where $\Delta \delta$ is the difference in the spectral power-law indices below and above the break. 

Meanwhile, it is difficult to imagine a compelling physical reason for the repeated change in the diffusion coefficient, now at 10 TV, and so another explanation is needed. 

\subsection{A local SNR surrounded by gas clouds} \label{LocalSNR}

There are many models discussing the origin of the observed spectral break at 200 GV \cite{2023FrPhy..1844301M}, but only a few of them are trying to reproduce the entire TV bump with two breaks at 200 GV and 10 TV. One of the most popular is the model speculating on the idea of the local SNR (e.g., Geminga SNR at $\sim$300 pc), surrounded by gas cloud(s) where the secondary species are produced. It is claimed that this model reproduces all the observed features in the spectra of CR nuclei, $e^\pm$, $\bar{p}$, and dipole anisotropy. Various versions of this model are discussed in \cite{2019PhRvD.100f3020Y, 2019JCAP...10..010L, 2020ApJ...903...69F, 2021PhRvD.104j3013F, 2021FrPhy..1624501Y, 2021PhRvD.104l3001Z, 2022ApJ...926...41Z, 2022ApJ...930...82L, 2023Galax..11...43L, 2022arXiv221205641Q, 2023ApJ...942...13Q, 2022MNRAS.511.6218Z, 2023JCAP...02..007Z, 2023ApJ...952..100N} and elsewhere.

In such a model \cite{2022arXiv221205641Q}, the local SNR accelerates primary species ($e^-$ and nuclei). Secondary species (LiBeB, $\bar{p}$, $e^\pm$) are produced by accelerated nuclei in the gas cloud(s) surrounding the shell. 
The nuclear species coming from the SNR have a \emph{convex} spectral shape, when scaled with $E_{\rm kin}^{2.6}$, formed by the cutoff of the injection spectrum $\propto$$R^{-\nu}e^{-R/R_c}$, with $R_c$$=$15 TV, while the low-energy 
decrease is formed due to a time delay in the propagation of particles from the SNR that have not yet reached us.
The SNR age, distance, and the diffusion coefficient are tuned in such a way that these bumps fit in between 200 GV and 100 TV. To match the observed CR spectra and make room for the SNR component, the spectra of Galactic CR species must have a \emph{concave} shape. The latter is done by adjusting the parameters of the two-halo scenario \cite{2015PhRvD..92h1301T}. The steepening in the nuclear spectra, which becomes visible at 5 TV, is tuned to reproduce the observed steepening in the $e^+$ spectrum at $\approx$300 GV, assuming that all observed excess positrons are produced in the proposed scenario (5\,TV/300\,GV$\approx$17 is the mean fraction of the kinetic energy of the primary proton transferred to the secondary pion per collision \cite{2000A&A...362..937A, 2006PhRvD..74c4018K}). Primary electrons lose energy due to the inverse Compton scattering and synchrotron emission to make the observed break at 1 TV (see Sect.\ \ref{CRe}). The suggested source is Geminga SNR with an age 330 kyr and a distance of 330 pc.


There are some issues with the local SNR model. First, it requires fine tuning. To fit the observed data, it is necessary to make a dip in the spectra of Galactic species and a bump in the corresponding local SNR components at the same energy simultaneously. The number of free parameters in this model, not counting the normalizations of the Galactic CR species, is about 50. These include 8 transport parameters + 6 spectral parameters + 28 (may be somewhat less) individual normalizations of SNR components for each species + 7 parameters for primary electrons + 1 gas cloud grammage. 

Second, a simple estimate of the diffusion length shows that it cannot reproduce the observed sharp jump in the relative CR intensity across the magnetic equator \cite{2019ApJ...871...96A}. Gyroradius of 10 TV particles in the interstellar 3 $\mu$G magnetic field is $\sim$0.003 pc. For a source at $\sim$330 pc, this gives $\sim$$10^5$ mean free paths -- there is no way to produce the sharp jump in the anisotropy map at such a distance. Even if the mean free path is $\sim$1 pc, it is still $\sim$330 mean free paths. In fact, all models of the TV bump with relatively distant sources have the same problem. 

This leads us to the conclusion that the ``source'' should be much closer to the solar system, and it should be a different type of source.

\subsection{Reacceleration bump}

The large number of free parameters discussed above can be avoided
if, instead of a source surrounded by gas clouds, we assume that pre-existing CRs are \emph{reaccelerated} in a local shock \cite{2021ApJ...911..151M, 2022ApJ...933...78M, Malkov:2023fQ}. If the shock is located at a few particle's mean free paths from the observer or connected to the observer by the magnetic field line, it can also explain the observed sharp jump in the relative CR intensity across the magnetic equator. Such a model requires only a moderate reacceleration below rigidity $\sim$50 TV, a shock with the Mach number of $\sim$1.5 should suffice. Reaccelerated particles below $\sim$0.2 TV are convected with the ISM flow and do not reach us, thus creating the bump. This single universal process acts on all CR species in the rigidity range below 100 TV. The position of the sun relative to the shock defines the bump parameters, which can change over time.

The model does not specify the location of the shock. The passing stars' bow shock, shock from the old SNR, or any local shock with a small Mach number $\sim$1.5 can do the job. The distance-shock-size relation gives an estimate of the distance: $\zeta_{obs}{\rm (pc)}$$\sim$$100\sqrt{L_\perp{\rm (pc)}}$; for sufficiently large bow shocks, $L_\perp$$=$$10^{-3}$--$10^{-2}$ pc, the distance is $\zeta_{obs}$$=$3--10 pc. 

The \emph{only three} unknown bump-specific fitting parameters can be obtained from a fit to the proton spectrum, best measured among CR species. Spectra of all other CR species can be calculated using a simple analytical formula, where the input parameters are their normalizations and spectral power-law indices derived from their local interstellar spectra (LIS) below the bump. The steeper the spectrum of ambient particles, the stronger the effect of reacceleration, and larger the bump. The LIS for all species H--Ni are provided in \cite{2020ApJS..250...27B}, and updated Fe LIS -- in \cite{2021ApJ...913....5B}. 

Interestingly that the increased CR intensity feature in the small-scale residual anisotropy map aligns well with the direction of the local magnetic field \cite{2019ApJ...871...96A} and may indicate the direction to the shock. One of the favorite candidates is the $\epsilon$ Eri star \cite{Malkov:2023fQ}, with its configuration of the bow and termination shocks, projected just 6.7$^\circ$ off of the direction of the local magnetic field. $\epsilon$ Eri is a solar like star, K2 dwarf (5 000 K) with the mass 0.82$M_\odot$ and radius 0.74$R_\odot$. It is located at the distance of 3.2 pc and has the speed 20 km/s, a bit small, but has an astonishing mass loss rate: $\dot{m}$$\sim$30--1500$\dot{M}_\odot$. It also has a huge astrosphere -- 8000 au, 47$'$ as seen from Earth (larger than the angular size of the Moon!).

Other candidate stars are $\epsilon$ Indi, a triplet star K4.5V (0.77$M_\odot$)+T1.5 (0.072$M_\odot$)+T6 (0.067$M_\odot$) at a distance of 3.6 pc and moving with the speed of 40.4 km/s (radial), or Scholz's Star, a duplet M9.5 (0.095$M_\odot$)+T5.5 (0.063$M_\odot$) at a distance of 6.8 pc, moving with the speed of 82.4 km/s (radial). There could be other types of shocks in the solar neighborhood, see discussion in \cite{2021ApJ...911..151M, Malkov:2023fQ}.

\section{Cosmic ray electrons} \label{CRe}

Electrons in CRs are subject to severe energy losses at all energies. The fastest losses are at low and high energies due to the ionization and the inverse Compton and synchrotron emission, correspondingly. Therefore, in order for very-high energy electrons to reach us, their sources must be close to us and relatively young. Perhaps Nishimura et al. \cite{1979ICRC....1..488N} were the first to suggest that ``the electron spectrum in TeV region would deviate from smooth power law behavior due to small number of sources which are capable of contributing to the observed flux... several bumps would be observed in the spectrum correlating to each source...'' Other early papers \cite{1995A&A...294L..41A, 2004ApJ...601..340K} also show possible contributions of local sources. Subsequent publications discussed the origin of the observed spectrum and modeled the contribution of local sources above 1 TeV.

The all-electron ($e^-$$+$$e^+$) spectrum up to 1 TeV was measured by \fermilat{} \cite{2009PhRvL.102r1101A, 2010PhRvD..82i2004A} (pre-\fermilat{} measurements are summarized in \cite{2009PhRvL.102r1101A}). It appears to be too flat, contrary to the expectations of a steep decrease (see Fig.~57 in \cite{2021PhR...894....1A} and Fig.~5 in \cite{Adriani:2023SR}), and cannot be reproduced with a single component. The sharp cutoff above $\sim$1 TeV was reported by H.E.S.S.\  \cite{2008PhRvL.101z1104A, 2009A&A...508..561A} and confirmed by VERITAS \cite{2018PhRvD..98f2004A}. Subsequent measurements by PAMELA \cite{2017RNCim.40...473A},  AMS-02 \cite{2014PhRvL.113v1102A, 2019PhRvL.122j1101A, 2021PhR...894....1A}, CALET \cite{2018PhRvL.120z1102A}, DAMPE \cite{2017Natur.552...63D}, and \fermilat\ \cite{2017PhRvD..95h2007A} 
reveal a flattening at $\sim$60 GeV and confirmed the cutoff at 1 TeV. Interestingly, the experiments are consistent in pairs, that is, CALET and AMS-02 are consistent with each other, but differ by $\approx$20\% from DAMPE and \fermilat{}, which are also consistent with each other. Above 1 TeV, H.E.S.S., CALET, and DAMPE measurements are consistent with each other, there is also a slight hint at a possible feature above 3 TeV albeit with large error bars. Slow diffusion zones observed around several pulsars \cite{2017Sci...358..911A} can increase the energy losses of TeV electrons making observations of such features less likely.
The all-electron spectrum includes the $e^+$ excess (the so-called ``signal''), and may include an identical $e^-$ ``signal'' if the source is charge-sign symmetric, such as pulsars or dark matter (DM) \cite{2021PhR...894....1A}. 

There are quite a number of publications discussing the contributions to the all-electron spectrum from local sources. Such multi-component models include the average electron spectrum from distant Galactic sources, parametrized contributions from local catalog sources (SNRs, PWNe), and may utilize the observed radio spectral indices of local SNRs, see examples in \cite{2017JCAP...01..006M, Motz:2023AX}. 

AMS-02 data on electrons ($e^-$) may offer a clue to the origin of the break at $\sim$1 TeV. Preliminary analysis \cite{Kounine:2023AV} uses a three-component fit, which includes low- and high-energy power-laws plus the $e^+$-like source term. According to this fit, the break in the all-electron spectrum at $\sim$1 TeV is related to the cutoff in the $e^+$ spectrum plus the identical $e^-$ component, and implies a charge-symmetric source of the excess positrons (pulsars, DM). However, more accurate data is needed to test if the charge-symmetry is exact (e.g., hadronic processes do not produce identical $e^\pm$ spectra).

\section{Cosmic ray positrons}

Unexpected behavior of the positron fraction \pfrac{} was first noticed in the data from the TS93 balloon flight \cite{1996ApJ...457L.103G}, which observed a constant positron fraction of $0.078$$\pm$$0.016$ in the range of 5--60 GeV, and in the HEAT experiment \cite{2004PhRvL..93x1102B}, which detected ``a small positron flux of nonstandard origin.'' 
Earlier experiments were contaminated with protons due to insufficient rejection capability. The results did not attract much attention until PAMELA team reported a surprising rise in the positron fraction up to 100 GeV \cite{2009Natur.458..607A}, contrary to the expectations of a monotonic decrease with energy \cite{1982ApJ...254..391P, 1998ApJ...493..694M}. Conventional models imply a smooth CR source distribution and steady-state production of secondary species in the hadronic CR interactions with ISM gas. The rise was confirmed by \fermilat{} \cite{2012PhRvL.108a1103A}, which used the geomagnetic field for identifying positrons, and with a higher precision and up to $\sim$500 GeV by AMS-02 \cite{2014PhRvL.113l1101A, 2019PhRvL.122d1102A}. 
The latest AMS-02 data indicate that the $e^+$ flux is the sum of low-energy secondaries from CR 
production 
plus a high-energy component from a new source with a cutoff at $ 749^{+197}_{-137}$ GeV \cite{2021PhR...894....1A, Kounine:2023AV}.

Perhaps the most striking is the fact that the $e^+/\bar{p}$ and $\bar{p}/p$ ratios barely change from 60--525 GeV \cite{2016PhRvL.117i1103A, 2021PhR...894....1A} hinting at some connection between the three species. Fitting a constant to the flux ratio in this range yields $e^+/\bar{p}$\,$=$\,$2.01$\,$\pm$0.03(stat.)\,$\pm$0.05(syst.)\ \cite{Weng:20231Q}, consistent with a constant. It is unclear whether it has a fundamental importance or is a chance coincidence. Some authors argue that this indicates that both species ($e^+,\bar{p}$) are secondary and are produced in the same process \cite{2017PhRvD..95f3009L, 2019PhRvD..99d3005L}. However, the fit could also be performed with a moderately rising functional dependence, because constraining are only a few lower-energy points from 60--130 GeV. 

Given the high collected positron statistics, the authors \cite{2023ApJ...950..120C} point to small but significant peaks in the positron fraction at 12 and 21 GeV, which could be associated with a powerful explosion in the Galactic center and {\it Fermi} Bubbles \cite{2010ApJ...724.1044S, 2014ApJ...793...64A} or with yet unknown processes. 

The positron anomaly gave rise to a huge number of interpretation papers, most of which connect the positron excess with DM. However, accurate multi-messenger data ($\gamma, \bar{p}$, CRs) collected after the PAMELA discovery imposes tight constraints on the most simplistic DM models (see a review talk by Francesca Calore \cite{Calore:2023AV}).
The astrophysical interpretations can be divided into groups of models with primary and secondary origin of positrons. In turn, models with primary positrons discuss their production in pulsars with additional effects produced by pulsar bow shocks and slow diffusion zones. Secondary production models are divided between positron production in ISM and in SNR shocks; the latter exploit various properties of SNR shocks and various configurations of target gas distributions. There are also models exploiting inhomogeneity of the SNR (CR sources) distribution. Here we mention only a few examples.

\emph{A model with local SNR, surrounded by gas clouds (secondary production in the local ISM), has already been discussed above. We refer the reader to Sect.\ \ref{LocalSNR}.}

\subsection{Positron production in Galactic SNR shocks}

The first models \cite{2009PhRvL.103e1104B} speculated on the idea of producing secondary species in a SNR shock, where CRs are accelerated (originally proposed in \cite{2003A&A...410..189B}). Therefore, the secondary $e^\pm$ participate in the acceleration process and turn out to have a very flat spectrum, which is responsible, after propagation in the Galaxy, for the observed positron excess. However, it soon becomes clear that this process should also increase the production of other secondary species, and thus other secondary to primary ratios ($\bar{p}/p$, B/C) should rise too \cite{2009PhRvL.103h1103B, 2009PhRvL.103h1104M, 2013PhRvD..87d7301K, 2014PhRvD..89d3013C}, contrary to observations. Non-observation of such rise provides significant constraints on physical conditions in the shock. 


\subsection{Secondary positrons in the volume charge model}

In this model \cite{2016PhRvD..94f3006M}, CRs accelerated in the SNR shell penetrate into the dense gas clumps upstream, where they interact with the ISM gas and produce secondary particles ($\bar{p}, e^\pm$). The predominance of positively charged particles in the shock and in the pre-cursor develops a positive electric volume charge in the gas cloud, which preferentially expels secondary positrons into the upstream plasma where they are accelerated by the shock. Since the shock is modified, these positrons develop a harder spectrum than CR electrons accelerated in other SNRs. Mixing these populations explains the increase in the positron fraction \pfrac{} above 8 GeV. Besides, there are also other sources of positrons in the ISM, such as radioactive decay.

\subsection{Primary positrons from pulsars}

Pulsars are the primary charge-symmetric suspects as they disconnect positrons from nuclear species and therefore remove the constraints associated with production of other secondaries. Probably, the first mention of pulsars as sources of CR positrons can be found in  \cite{1981IAUS...94..175A}, and the first calculation of secondary production and pulsar contribution to CR positrons -- in \cite{1987ICRC....2...92H}. 
A calculation of the positron fraction using the data available at that time (contaminated by protons above $\approx$5 GeV) was performed in \cite{1989ApJ...342..807B}.
This model included secondary $e^\pm$, primary $e^-$ from SNR, and primary $e^\pm$ from pulsars. 
Examples of modern calculations of the positron fraction, which include $e^\pm$ from known sources and secondary $e^\pm$ can be found in \cite{2013ApJ...772...18L, 2017PhRvD..96j3013H}.

\subsection{Primary positrons in the pulsar bow shock model}

Pulsars with high spin-down power produce relativistic winds. Some pulsars move relative to their surrounding ISM at supersonic speeds producing bow shocks. Ultrarelativistic particles accelerated at the termination surface of the pulsar wind can experience reacceleration in the converging flow system, producing a universal spectrum similar to that of protons accelerated in the SNR shell. This scenario naturally explains why the $e^+/p$ ratio remains constant above 60 GeV. 
Primary positrons and electrons in this scenario have similar spectra. 

An idea of positron reacceleration in a pulsar bow shock was proposed in \cite{2017SSRv..207..235B}, and further detailed in \cite{2019ApJ...876L...8B}. It is suggested that the 5.7 millisecond (MSP) pulsar PSR J0437-4715 may produce the observed positrons. The pulsar distance and velocity are $\approx$$156.79$$\pm$$0.25$ pc and $\sim$100 km/s. It is the closest and brightest MSP in a binary system with a white dwarf companion and an orbital period of 5.7 days. 
It is observed in optical, far-ultraviolet (FUV), and X-ray bands and exhibits the greatest long-term rotational stability of any pulsar. The model assumes that the pulsar's position coincides with the direction of the local magnetic field and adjusts the parallel diffusion coefficient to match optical, FUV, and X-ray constraints on the flux of accelerated leptons from the nebula.

\section{Cosmic ray antiprotons and 10 GeV excess}

CR antiprotons in the range 2--12 GeV were observed for the first time during balloon flights \cite{1979PhRvL..43.1196G, 1979ICRC....1..330B}. The following series of Antarctic flights by BESS \cite{2013NuPhS.243...92Y}, N.\ America flights by MASS91 \cite{1996ApJ...467L..33H}, HEAT \cite{2001PhRvL..87A1101B}, and CAPRICE98 \cite{2001ApJ...561..787B} instruments, and space experiments, PAMELA \cite{2010PhRvL.105l1101A} and AMS-02 \cite{2021PhR...894....1A}, extended the energy range and increased accuracy of $\bar{p}$ measurements. The $\bar{p}$ spectrum is now measured up to $\sim$450 GV \cite{2021PhR...894....1A}. It features a low-energy rise caused by the kinematics of the process \cite{1998ApJ...499..250S}. Ratio $\bar{p}/p\approx const$ from 30--450 GV \cite{2021PhR...894....1A}.


Following the publication of the AMS-02 $\bar{p}$ data \cite{2016PhRvL.117i1103A}, several groups independently noticed an excess over conventional model predictions at around 10 GeV \cite{2017ApJ...840..115B, 2017PhRvL.118s1101C, 2017PhRvL.118s1102C}; all three papers are marked as published on May 9--12, 2017. Two papers \cite{2017PhRvL.118s1101C, 2017PhRvL.118s1102C} proposed an interpretation in terms of DM, while \cite{2017ApJ...840..115B} pointed to increased systematics due to the high solar activity period during data taking or due to the production cross section uncertaincies, see also \cite{2006ApJ...642..902P, 2020PhRvR...2d3017H, 2020PhRvD.102j3007E, 2021ApJ...908..167E, 2023arXiv230400760L}.

At present, the two hypotheses remain, (i) DM contribution, and (ii) systematics due to the solar modulation and/or cross section uncertainties. People like the DM hypothesis, but attempts are being made to improve on the cross sections. Interestingly, the same proposed DM candidate ($m_\chi\approx 50-100$ GeV) can reproduce the 10 GeV antiproton excess, $\gamma$-ray excess from thsce Galactic center, and the extended $\gamma$-ray emission from the 400-kpc-across halo of the Andromeda galaxy (M31) \cite{2019ApJ...880...95K, 2021PhRvD.103b3027K} (see also references therein).
Due to space limitations, I have to conclude my review with an incomplete list of papers that discuss astrophysical uncertainties associated with antiprotons \cite{2015ApJ...803...54K, 2023CoPhC.28708698K, 2017JCAP...02..048W, 2017PhRvD..96d3007D, 2018PhRvD..97j3019K, 2018PhRvL.121v2001A, 2020PhRvR...2b3022B}, and the DM interpretation of the antiproton excess \cite{2019PhRvD..99j3026C, 2021MPLA...3630003H, 2022ScPP...12..163C, 2022PhRvL.129w1101Z}.

\section{Concluding remarks -- situation with production cross sections}

Unfortunately, one major remaining bottleneck is the accuracy of particle and especially isotopic production cross sections (see discussion in \cite{2018PhRvC..98c4611G, 2023arXiv230706798G}). Every time an unexpected spectral feature is detected, there is a chance of an error in the model predictions due to errors in the cross sections. The elimination of such errors is easier and much cheaper than building and successfully launching a major space mission like AMS-02, but this requires a dedicated community effort. 

\smallskip
\emph{Partial support from NASA grants Nos.\ 80NSSC22K0477, 80NSSC22K0718, 80NSSC23K0169 is greatly acknowledged.}

\small
\bibliography{review}

\providecommand{\href}[2]{#2}\begingroup\raggedright\begin{thebibliography}{100}

\bibitem{2001SSRv...99...85S}
S.P.~{Swordy}, \emph{{The Energy Spectra and Anisotropies of Cosmic Rays}},
  \href{https://doi.org/10.1023/A:1013828611730}{\emph{\ssr} {\bfseries 99}
  (2001) 85}.

\bibitem{1990YadF...51...99G}
N.L.~{Grigorov}, \emph{{Study of cosmic rays of high and superhigh energy in
  satellites}}, {\emph{Soviet J. Nuclear Phys.} {\bfseries 51} (1990) 99}.

\bibitem{1993RoIzF..57...76I}
I.P.~{Ivanenko}, V.Y.~{Shestoperov}, D.M.~{Podorozhnyj} et~al., \emph{{Energy
  spectra of different cosmic-ray components at energies higher than 2 TeV
  measured by the SOKOL facility}}, {\emph{Bull.\ Russian Acad.\ of Sciences.
  Physics} {\bfseries 57} (1993) 76}.

\bibitem{1998ApJ...502..278A}
K.~{Asakimori}, T.H.~{Burnett}, M.L.~{Cherry} et~al., \emph{{Cosmic-Ray Proton
  and Helium Spectra: Results from the JACEE Experiment}},
  \href{https://doi.org/10.1086/305882}{\emph{\apj} {\bfseries 502} (1998)
  278}.

\bibitem{2019NIMPA.94462561S}
S.~{Schael}, A.~{Atanasyan}, J.~{Berdugo} et~al., \emph{{AMS-100: The next
  generation magnetic spectrometer in space - An international science platform
  for physics and astrophysics at Lagrange point 2}},
  \href{https://doi.org/10.1016/j.nima.2019.162561}{\emph{\nima} {\bfseries
  944} (2019) 162561}.

\bibitem{2023EPJWC.28001008P}
C.~{Perrina}, \emph{{Attaining the PeV frontier of the cosmic ray spectrum in
  space with HERD}}, {\emph{\epjwc} {\bfseries 280} (2023) 01008}.

\bibitem{2019AdSpR..64.2619K}
D.~{Karmanov}, A.~{Panov}, D.~{Podorozhny} et~al., \emph{{The HERO (High Energy
  Ray Observatory) detector current status}},
  \href{https://doi.org/10.1016/j.asr.2019.04.025}{\emph{\asr} {\bfseries 64}
  (2019) 2619}.

\bibitem{2021ExA....51.1299B}
R.~{Battiston}, B.~{Bertucci}, O.~{Adriani} et~al., \emph{{High precision
  particle astrophysics as a new window on the universe with an Antimatter
  Large Acceptance Detector In Orbit (ALADInO)}},
  \href{https://doi.org/10.1007/s10686-021-09708-w}{\emph{Exp.\ Astron.}
  {\bfseries 51} (2021) 1299}.

\bibitem{1990acr..book.....B}
V.S.~{Berezinskii}, S.V.~{Bulanov}, V.A.~{Dogiel} et~al., \emph{{Astrophysics
  of cosmic rays}}. {{Ginzburg}, V.~L., ed., Amsterdam: North-Holland, 1990}.

\bibitem{1999PhyU...42..353G}
V.L.~{Ginzburg}, \emph{{PHYSICS OF OUR DAYS: What problems of physics and
  astrophysics seem now to be especially important and interesting (thirty
  years later, already on the verge of XXI century)?}},
  \href{https://doi.org/10.1070/PU1999v042n04ABEH000562}{\emph{Physics Uspekhi}
  {\bfseries 42} (1999) 353}.

\bibitem{2016ApJ...831...18C}
A.C.~{Cummings}, E.C.~{Stone}, B.C.~{Heikkila} et~al., \emph{{Galactic Cosmic
  Rays in the Local Interstellar Medium: Voyager 1 Observations and Model
  Results}}, \href{https://doi.org/10.3847/0004-637X/831/1/18}{\emph{\apj}
  {\bfseries 831} (2016) 18}.

\bibitem{2009PhRvL.103g2502R}
G.~{Rugel}, T.~{Faestermann}, K.~{Knie} et~al., \emph{{New Measurement of the
  $^{60}$Fe Half-Life}},
  \href{https://doi.org/10.1103/PhysRevLett.103.072502}{\emph{\prl} {\bfseries
  103} (2009) 072502}.

\bibitem{2015ApJ...800...71F}
B.J.~{Fry}, B.D.~{Fields}, J.R.~{Ellis}, \emph{{Astrophysical Shrapnel:
  Discriminating among Near-Earth Stellar Explosion Sources of Live Radioactive
  Isotopes}}, \href{https://doi.org/10.1088/0004-637X/800/1/71}{\emph{\apj}
  {\bfseries 800} (2015) 71}.

\bibitem{2016Natur.532...69W}
A.~{Wallner}, J.~{Feige}, N.~{Kinoshita} et~al., \emph{{Recent near-Earth
  supernovae probed by global deposition of interstellar radioactive
  $^{60}$Fe}}, \href{https://doi.org/10.1038/nature17196}{\emph{\nat}
  {\bfseries 532} (2016) 69}.

\bibitem{2016Natur.532...73B}
D.~{Breitschwerdt}, J.~{Feige}, M.M.~{Schulreich} et~al., \emph{{The locations
  of recent supernovae near the Sun from modelling $^{60}$Fe transport}},
  \href{https://doi.org/10.1038/nature17424}{\emph{\nat} {\bfseries 532} (2016)
  73}.

\bibitem{1999A&A...346..785S}
D.M.~{Sfeir}, R.~{Lallement}, F.~{Crifo} et~al., \emph{{Mapping the contours of
  the Local bubble: preliminary results}}, {\emph{\aap} {\bfseries 346} (1999)
  785}.

\bibitem{2011ARA&A..49..237F}
P.C.~{Frisch}, S.~{Redfield}, J.D.~{Slavin}, \emph{{The Interstellar Medium
  Surrounding the Sun}},
  \href{https://doi.org/10.1146/annurev-astro-081710-102613}{\emph{\araa}
  {\bfseries 49} (2011) 237}.

\bibitem{1999PhRvL..83...18K}
K.~{Knie}, G.~{Korschinek}, T.~{Faestermann} et~al., \emph{{Indication for
  Supernova Produced $^{60}$Fe Activity on Earth}},
  \href{https://doi.org/10.1103/PhysRevLett.83.18}{\emph{\prl} {\bfseries 83}
  (1999) 18}.

\bibitem{2004PhRvL..93q1103K}
K.~{Knie}, G.~{Korschinek}, T.~{Faestermann} et~al., \emph{{$^{60}$Fe Anomaly
  in a Deep-Sea Manganese Crust and Implications for a Nearby Supernova
  Source}}, \href{https://doi.org/10.1103/PhysRevLett.93.171103}{\emph{\prl}
  {\bfseries 93} (2004) 171103}.

\bibitem{2016PNAS..113.9232L}
P.~{Ludwig}, S.~{Bishop}, R.~{Egli} et~al., \emph{{Time-resolved
  2-million-year-old supernova activity discovered in Earth's microfossil
  record}}, \href{https://doi.org/10.1073/pnas.1601040113}{\emph{Proc.\ Nat.\
  Acad.\ of Science} {\bfseries 113} (2016) 9232}.

\bibitem{2009LPI....40.1129C}
D.L.~{Cook}, E.~{Berger}, T.~{Faestermann} et~al., \emph{{$^{60}$Fe, $^{10}$Be,
  and $^{26}$Al in Lunar Cores 12025/8 and 60006/7: Search for a Nearby
  Supernova}},  in \emph{40th Annual Lunar and Planetary Science Conference},
  2009, p.1129.

\bibitem{2012LPI....43.1279F}
L.~{Fimiani}, D.L.~{Cook}, T.~{Faestermann} et~al., \emph{{Sources of Live
  $^{60}$Fe, $^{10}$Be, and $^{26}$Al in Lunar Core 12025, Core 15008, Skim
  Sample 69921, Scoop Sample 69941, and Under-Boulder Sample 69961}},  in
  \emph{43rd Annual Lunar and Planetary Science Conference}, 2012, p.1279.

\bibitem{2014LPI....45.1778F}
L.~{Fimiani}, D.L.~{Cook}, T.~{Faestermann} et~al., \emph{{Evidence for
  Deposition of Interstellar Material on the Lunar Surface}},  in \emph{45th
  Annual Lunar and Planetary Science Conference}, 2014, p.1778.

\bibitem{2019PhRvL.123g2701K}
D.~{Koll}, G.~{Korschinek}, T.~{Faestermann} et~al., \emph{{Interstellar
  $^{60}$Fe in Antarctica}},
  \href{https://doi.org/10.1103/PhysRevLett.123.072701}{\emph{\prl} {\bfseries
  123} (2019) 072701}.

\bibitem{2016Sci...352..677B}
W.R.~{Binns}, M.H.~{Israel}, E.R.~{Christian} et~al., \emph{{Observation of the
  $^{60}$Fe nucleosynthesis-clock isotope in galactic cosmic rays}},
  \href{https://doi.org/10.1126/science.aad6004}{\emph{Science} {\bfseries 352}
  (2016) 677}.

\bibitem{1999ApJ...523L..61W}
M.E.~{Wiedenbeck}, W.R.~{Binns}, E.R.~{Christian} et~al., \emph{{Constraints on
  the Time Delay between Nucleosynthesis and Cosmic-Ray Acceleration from
  Observations of $^{59}$Ni and $^{59}$Co}},
  \href{https://doi.org/10.1086/312242}{\emph{\apjl} {\bfseries 523} (1999)
  L61}.

\bibitem{2021ApJ...913....5B}
M.J.~{Boschini}, S.~{Della Torre}, M.~{Gervasi} et~al., \emph{{The Discovery of
  a Low-energy Excess in Cosmic-Ray Iron: Evidence of the Past Supernova
  Activity in the Local Bubble}},
  \href{https://doi.org/10.3847/1538-4357/abf11c}{\emph{\apj} {\bfseries 913}
  (2021) 5}.

\bibitem{2021PhRvL.126d1104A}
M.~{Aguilar}, L.A.~{Cavasonza}, M.S.~{Allen} et~al., \emph{{Properties of Iron
  Primary Cosmic Rays: Results from the Alpha Magnetic Spectrometer}},
  \href{https://doi.org/10.1103/PhysRevLett.126.041104}{\emph{\prl} {\bfseries
  126} (2021) 041104}.

\bibitem{2019AdSpR..64.2459B}
M.J.~{Boschini}, S.~{Della Torre}, M.~{Gervasi} et~al., \emph{{The HELMOD model
  in the works for inner and outer heliosphere: From AMS to Voyager probes
  observations}},
  \href{https://doi.org/10.1016/j.asr.2019.04.007}{\emph{Advances in Space
  Research} {\bfseries 64} (2019) 2459}.

\bibitem{2022PhRvL.128m1103A}
O.~{Adriani}, Y.~{Akaike}, K.~{Asano} et~al., \emph{{Direct Measurement of the
  Nickel Spectrum in Cosmic Rays in the Energy Range from 8.8 GeV/n to 240
  GeV/n with CALET on the International Space Station}},
  \href{https://doi.org/10.1103/PhysRevLett.128.131103}{\emph{\prl} {\bfseries
  128} (2022) 131103}.

\bibitem{Torii:20231Q}
S.~Torii, \emph{{Highlights from the CALET observations for 7.5 years on the
  International Space Station}},  in \emph{\ICRCthirtyeight {\textemdash}
  PoS(ICRC2023)}, vol.~444, 2023, 002.

\bibitem{2021PhRvL.127b1101A}
M.~{Aguilar}, L.A.~{Cavasonza}, B.~{Alpat} et~al., \emph{{Properties of a New
  Group of Cosmic Nuclei: Results from the Alpha Magnetic Spectrometer on
  Sodium, Aluminum, and Nitrogen}},
  \href{https://doi.org/10.1103/PhysRevLett.127.021101}{\emph{\prl} {\bfseries
  127} (2021) 021101}.

\bibitem{2022ApJ...933..147B}
M.J.~{Boschini}, S.D.~{Torre}, M.~{Gervasi} et~al., \emph{{Spectra of
  Cosmic-Ray Sodium and Aluminum and Unexpected Aluminum Excess}},
  \href{https://doi.org/10.3847/1538-4357/ac7443}{\emph{\apj} {\bfseries 933}
  (2022) 147}.

\bibitem{2020ApJS..250...27B}
M.J.~{Boschini}, S.~{Della Torre}, M.~{Gervasi} et~al., \emph{{Inference of the
  Local Interstellar Spectra of Cosmic-Ray Nuclei Z {\ensuremath{\leq}} 28 with
  the GALPROP-HELMOD Framework}},
  \href{https://doi.org/10.3847/1538-4365/aba901}{\emph{\apjs} {\bfseries 250}
  (2020) 27}.

\bibitem{BAUHOFF1986429}
W.~Bauhoff, \emph{Tables of reaction and total cross sections for
  proton-nucleus scattering below 1 gev},
  \href{https://doi.org/https://doi.org/10.1016/0092-640X(86)90016-1}{\emph{Atomic
  Data and Nuclear Data Tables} {\bfseries 35} (1986) 429}.

\bibitem{Bobchenko:1979hp}
B.M.~Bobchenko et~al., \emph{{Measurement of total inelastic cross-sections
  from proton interactions with nuclei in the momentum range from 5-GeV/c to
  9-GeV/c and pi-mesons with nuclei in the momentum range from 1.75-GeV/c to
  6.5-GeV/c}}, {\emph{Sov. J. Nucl. Phys.} {\bfseries 30} (1979) 805}.

\bibitem{2013ICRC...33..803M}
I.V.~{Moskalenko}, A.E.~{Vladimirov}, T.A.~{Porter} et~al., \emph{{Isotopic
  Production Cross Sections (ISOPROCS Project)}}, {\emph{33rd \icrc\ (Rio de
  Janeiro)} (2013) 0823}.

\bibitem{1991ICRC....2..280H}
S.E.~{Hirzebruch}, W.~{Heinrich}, K.D.~{Tolstov} et~al., \emph{{Nuclear
  Fragmentation Cross Sections at Relativistic Energies}},  in
  \emph{\ICRCtwentytwo}, vol.~2, 1991, p.280.

\bibitem{1999A&A...345..813K}
J.~{Kn{\"o}dlseder}, D.~{Dixon}, K.~{Bennett} et~al., \emph{{Image
  reconstruction of COMPTEL 1.8 MeV $^{26}$Al line data}}, {\emph{\aap}
  {\bfseries 345} (1999) 813}.

\bibitem{2015ApJ...801..142B}
L.~{Bouchet}, E.~{Jourdain}, J.-P.~{Roques}, \emph{{The Galactic $^{26}$Al
  Emission Map as Revealed by INTEGRAL SPI}},
  \href{https://doi.org/10.1088/0004-637X/801/2/142}{\emph{\apj} {\bfseries
  801} (2015) 142}.

\bibitem{1996PhR...267....1P}
N.~{Prantzos}, R.~{Diehl}, \emph{{Radioactive $^{26}$Al in the galaxy:
  observations versus theory}},
  \href{https://doi.org/10.1016/0370-1573(95)00055-0}{\emph{\physrep}
  {\bfseries 267} (1996) 1}.

\bibitem{2006Natur.439...45D}
R.~{Diehl}, H.~{Halloin}, K.~{Kretschmer} et~al., \emph{{Radioactive $^{26}$Al
  from massive stars in the Galaxy}},
  \href{https://doi.org/10.1038/nature04364}{\emph{\nat} {\bfseries 439} (2006)
  45}.

\bibitem{2002ApJ...576L..41M}
P.R.~{McCullough}, B.D.~{Fields}, V.~{Pavlidou}, \emph{{Discovery of an Old,
  Nearby, and Overlooked Supernova Remnant Centered on the Southern
  Constellation Antlia Pneumatica}},
  \href{https://doi.org/10.1086/343100}{\emph{\apjl} {\bfseries 576} (2002)
  L41}.

\bibitem{2022ApJ...940...63R}
S.~{Ranasinghe}, D.~{Leahy}, \emph{{Distances, Radial Distribution, and Total
  Number of Galactic Supernova Remnants}},
  \href{https://doi.org/10.3847/1538-4357/ac940a}{\emph{\apj} {\bfseries 940}
  (2022) 63}.

\bibitem{2020ApJ...889..167B}
M.J.~{Boschini}, S.~{Della Torre}, M.~{Gervasi} et~al., \emph{{Deciphering the
  Local Interstellar Spectra of Secondary Nuclei with the Galprop/Helmod
  Framework and a Hint for Primary Lithium in Cosmic Rays}},
  \href{https://doi.org/10.3847/1538-4357/ab64f1}{\emph{\apj} {\bfseries 889}
  (2020) 167}.

\bibitem{2013ApJ...770..117L}
K.A.~{Lave}, M.E.~{Wiedenbeck}, W.R.~{Binns} et~al., \emph{{Galactic Cosmic-Ray
  Energy Spectra and Composition during the 2009-2010 Solar Minimum Period}},
  \href{https://doi.org/10.1088/0004-637X/770/2/117}{\emph{\apj} {\bfseries
  770} (2013) 117}.

\bibitem{2018PhRvL.120b1101A}
M.~{Aguilar}, L.~{Ali Cavasonza}, G.~{Ambrosi} et~al., \emph{{Observation of
  New Properties of Secondary Cosmic Rays Lithium, Beryllium, and Boron by the
  Alpha Magnetic Spectrometer on the International Space Station}},
  \href{https://doi.org/10.1103/PhysRevLett.120.021101}{\emph{\prl} {\bfseries
  120} (2018) 021101}.

\bibitem{2018PhRvC..98c4611G}
Y.~{G{\'e}nolini}, D.~{Maurin}, I.V.~{Moskalenko} et~al., \emph{{Current status
  and desired precision of the isotopic production cross sections relevant to
  astrophysics of cosmic rays: Li, Be, B, C, and N}},
  \href{https://doi.org/10.1103/PhysRevC.98.034611}{\emph{\prc} {\bfseries 98}
  (2018) 034611}.

\bibitem{2020A&A...639A.131W}
N.~{Weinrich}, Y.~{G{\'e}nolini}, M.~{Boudaud} et~al., \emph{{Combined analysis
  of AMS-02 (Li,Be,B)/C, N/O, $^{3}$He, and $^{4}$He data}},
  \href{https://doi.org/10.1051/0004-6361/202037875}{\emph{\aap} {\bfseries
  639} (2020) A131}.

\bibitem{2021JCAP...07..010D}
P.~{De La Torre Luque}, M.N.~{Mazziotta}, F.~{Loparco} et~al., \emph{{Markov
  chain Monte Carlo analyses of the flux ratios of B, Be and Li with the
  DRAGON2 code}},
  \href{https://doi.org/10.1088/1475-7516/2021/07/010}{\emph{\jcap} {\bfseries
  2021} (2021) 010}.

\bibitem{2022A&A...668A...7M}
D.~{Maurin}, E.~{Ferronato Bueno}, Y.~{G{\'e}nolini} et~al., \emph{{The
  importance of Fe fragmentation for LiBeB analyses. Is a Li primary source
  needed to explain AMS-02 data?}},
  \href{https://doi.org/10.1051/0004-6361/202243446}{\emph{\aap} {\bfseries
  668} (2022) A7}.

\bibitem{1955ApJ...121..144C}
A.G.W.~{Cameron}, \emph{{Origin of Anomalous Abundances of the Elements in
  Giant Stars.}}, \href{https://doi.org/10.1086/145970}{\emph{\apj} {\bfseries
  121} (1955) 144}.

\bibitem{1971ApJ...164..111C}
A.G.W.~{Cameron}, W.A.~{Fowler}, \emph{{Lithium and the s-process in Red-Giant
  Stars}}, \href{https://doi.org/10.1086/150821}{\emph{\apj} {\bfseries 164}
  (1971) 111}.

\bibitem{1975A&A....42...55A}
M.~{Arnould}, H.~{Norgaard}, \emph{{The Explosive Thermonuclear Formation of
  $^7$Li and $^{11}$B}}, {\emph{\aap} {\bfseries 42} (1975) 55}.

\bibitem{1978ApJ...222..600S}
S.~{Starrfield}, J.W.~{Truran}, W.M.~{Sparks} et~al., \emph{{On $^{7}$Li
  production in nova explosions}},
  \href{https://doi.org/10.1086/156175}{\emph{\apj} {\bfseries 222} (1978)
  600}.

\bibitem{1996ApJ...465L..27H}
M.~{Hernanz}, J.~{Jose}, A.~{Coc} et~al., \emph{{On the Synthesis of $^7$Li and
  $^7$Be in Novae}}, \href{https://doi.org/10.1086/310122}{\emph{\apjl}
  {\bfseries 465} (1996) L27}.

\bibitem{2015Natur.518..381T}
A.~{Tajitsu}, K.~{Sadakane}, H.~{Naito} et~al., \emph{{Explosive lithium
  production in the classical nova V339 Del (Nova Delphini 2013)}},
  \href{https://doi.org/10.1038/nature14161}{\emph{\nat} {\bfseries 518} (2015)
  381}.

\bibitem{2015Natur.518..307H}
M.~{Hernanz}, \emph{{Astrophysics: A lithium-rich stellar explosion}},
  \href{https://doi.org/10.1038/518307a}{\emph{\nat} {\bfseries 518} (2015)
  307}.

\bibitem{2018ApJ...862..141M}
W.~{Menn}, E.A.~{Bogomolov}, M.~{Simon} et~al., \emph{{Lithium and Beryllium
  Isotopes with the PAMELA Experiment}},
  \href{https://doi.org/10.3847/1538-4357/aacf89}{\emph{\apj} {\bfseries 862}
  (2018) 141}.

\bibitem{Wei:20231Q}
J.~Wei, \emph{{Cosmic-Ray Lithium and Beryllium Isotopes with the Alpha
  Magnetic Spectrometer}},  in \emph{\ICRCthirtyeight {\textemdash}
  PoS(ICRC2023)}, vol.~444, 2023, 077.

\bibitem{2011Sci...332...69A}
O.~{Adriani}, G.C.~{Barbarino}, G.A.~{Bazilevskaya} et~al., \emph{{PAMELA
  Measurements of Cosmic-Ray Proton and Helium Spectra}},
  \href{https://doi.org/10.1126/science.1199172}{\emph{Science} {\bfseries 332}
  (2011) 69}.

\bibitem{2021PhR...894....1A}
M.~{Aguilar}, L.~{Ali Cavasonza}, G.~{Ambrosi} et~al., \emph{{The Alpha
  Magnetic Spectrometer (AMS) on the international space station: Part II -
  Results from the first seven years}},
  \href{https://doi.org/10.1016/j.physrep.2020.09.003}{\emph{\physrep}
  {\bfseries 894} (2021) 1}.

\bibitem{Adriani:2023E5}
O.~Adriani, Y.~Akaike, K.~Asano et~al., \emph{{Helium flux and its ratio to
  proton flux in cosmic rays measured with CALET on the International Space
  Station}},  in \emph{\ICRCthirtyeight {\textemdash} PoS(ICRC2023)}, vol.~444,
  2023, 054.

\bibitem{Coppin:2023Px}
P.~Coppin, A.~Kotenko, P.-X.~Ma et~al., \emph{{Analysis of Individual
  Cosmic-Ray Proton and Helium Fluxes towards PeV Energies with DAMPE}},  in
  \emph{\ICRCthirtyeight {\textemdash} PoS(ICRC2023)}, vol.~444, 2023, 170.

\bibitem{1990ICRC....3...77I}
P.I.~{Ivanenko}, D.I.~{Rapoport}, Y.V.~{Shestoperov} et~al., \emph{{Energy
  Spectrum and Cosmic Ray Composition in the Region of Energies Higher than 1
  TeV Investigated Onboard the ``Cosmos-1543'' and ``Cosmos-1713''
  Satellites}},  in \emph{\ICRCtwentyone}, vol.~3, 1990, p.77.

\bibitem{2009BRASP..73..564P}
A.D.~{Panov}, J.H.~{Adams}, H.S.~{Ahn} et~al., \emph{{Energy spectra of
  abundant nuclei of primary cosmic rays from the data of ATIC-2 experiment:
  Final results}},
  \href{https://doi.org/10.3103/S1062873809050098}{\emph{Bull.\ Russian Acad.\
  Sci., Physics} {\bfseries 73} (2009) 564}.

\bibitem{2010ApJ...714L..89A}
H.S.~{Ahn}, P.~{Allison}, M.G.~{Bagliesi} et~al., \emph{{Discrepant Hardening
  Observed in Cosmic-ray Elemental Spectra}},
  \href{https://doi.org/10.1088/2041-8205/714/1/L89}{\emph{\apjl} {\bfseries
  714} (2010) L89}.

\bibitem{2011ApJ...729L..13O}
Y.~{Ohira}, K.~{Ioka}, \emph{{Cosmic-ray Helium Hardening}},
  \href{https://doi.org/10.1088/2041-8205/729/1/L13}{\emph{\apjl} {\bfseries
  729} (2011) L13}.

\bibitem{2016PhRvD..93h3001O}
Y.~{Ohira}, N.~{Kawanaka}, K.~{Ioka}, \emph{{Cosmic-ray hardenings in light of
  AMS-02 data}}, \href{https://doi.org/10.1103/PhysRevD.93.083001}{\emph{\prd}
  {\bfseries 93} (2016) 083001}.

\bibitem{2015ApJ...803L..15T}
N.~{Tomassetti}, F.~{Donato}, \emph{{The Connection between the Positron
  Fraction Anomaly and the Spectral Features in Galactic Cosmic-ray Hadrons}},
  \href{https://doi.org/10.1088/2041-8205/803/2/L15}{\emph{\apjl} {\bfseries
  803} (2015) L15}.

\bibitem{2019PhRvD.100f3020Y}
R.~{Yang}, F.~{Aharonian}, \emph{{Interpretation of the excess of antiparticles
  within a modified paradigm of galactic cosmic rays}},
  \href{https://doi.org/10.1103/PhysRevD.100.063020}{\emph{\prd} {\bfseries
  100} (2019) 063020}.

\bibitem{2019ApJ...872..108H}
A.~{Hanusch}, T.V.~{Liseykina}, M.~{Malkov}, \emph{{Acceleration of Cosmic Rays
  in Supernova Shocks: Elemental Selectivity of the Injection Mechanism}},
  \href{https://doi.org/10.3847/1538-4357/aafdae}{\emph{\apj} {\bfseries 872}
  (2019) 108}.

\bibitem{2017PhRvL.119q1101C}
D.~{Caprioli}, D.T.~{Yi}, A.~{Spitkovsky}, \emph{{Chemical Enhancements in
  Shock-Accelerated Particles: Ab initio Simulations}},
  \href{https://doi.org/10.1103/PhysRevLett.119.171101}{\emph{\prl} {\bfseries
  119} (2017) 171101}.

\bibitem{2021PhRvL.126h1102A}
M.~{Aguilar}, L.A.~{Cavasonza}, M.S.~{Allen} et~al., \emph{{Properties of Heavy
  Secondary Fluorine Cosmic Rays: Results from the Alpha Magnetic
  Spectrometer}},
  \href{https://doi.org/10.1103/PhysRevLett.126.081102}{\emph{\prl} {\bfseries
  126} (2021) 081102}.

\bibitem{2022ApJ...925..108B}
M.J.~{Boschini}, S.~{Della Torre}, M.~{Gervasi} et~al., \emph{{A Hint of a
  Low-energy Excess in Cosmic-Ray Fluorine}},
  \href{https://doi.org/10.3847/1538-4357/ac313d}{\emph{\apj} {\bfseries 925}
  (2022) 108}.

\bibitem{2023PhRvD.107f3020Z}
M.-J.~{Zhao}, X.-J.~{Bi}, K.~{Fang}, \emph{{Can the production cross-section
  uncertainties explain the cosmic fluorine anomaly?}},
  \href{https://doi.org/10.1103/PhysRevD.107.063020}{\emph{\prd} {\bfseries
  107} (2023) 063020}.

\bibitem{2014PhRvL.112o1103A}
M.~{Ackermann}, M.~{Ajello}, A.~{Albert} et~al., \emph{{Inferred Cosmic-Ray
  Spectrum from Fermi Large Area Telescope {\ensuremath{\gamma}}-Ray
  Observations of Earth's Limb}},
  \href{https://doi.org/10.1103/PhysRevLett.112.151103}{\emph{\prl} {\bfseries
  112} (2014) 151103}.

\bibitem{2017ApJ...839....5Y}
Y.S.~{Yoon}, T.~{Anderson}, A.~{Barrau} et~al., \emph{{Proton and Helium
  Spectra from the CREAM-III Flight}},
  \href{https://doi.org/10.3847/1538-4357/aa68e4}{\emph{\apj} {\bfseries 839}
  (2017) 5}.

\bibitem{2018JETPL.108....5A}
E.~{Atkin}, V.~{Bulatov}, V.~{Dorokhov} et~al., \emph{{New Universal Cosmic-Ray
  Knee near a Magnetic Rigidity of 10 TV with the NUCLEON Space Observatory}},
  \href{https://doi.org/10.1134/S0021364018130015}{\emph{JETP Letters}
  {\bfseries 108} (2018) 5}.

\bibitem{2019SciA....5.3793A}
Q.~{An}, R.~{Asfandiyarov}, P.~{Azzarello} et~al., \emph{{Measurement of the
  cosmic ray proton spectrum from 40 GeV to 100 TeV with the DAMPE satellite}},
  \href{https://doi.org/10.1126/sciadv.aax3793}{\emph{Science Advances}
  {\bfseries 5} (2019) eaax3793}.

\bibitem{2022PhRvL.129j1102A}
O.~{Adriani}, Y.~{Akaike}, K.~{Asano} et~al., \emph{{Observation of Spectral
  Structures in the Flux of Cosmic-Ray Protons from 50 GeV to 60 TeV with the
  Calorimetric Electron Telescope on the International Space Station}},
  \href{https://doi.org/10.1103/PhysRevLett.129.101102}{\emph{\prl} {\bfseries
  129} (2022) 101102}.

\bibitem{2022ApJ...940..107C}
G.H.~{Choi}, E.S.~{Seo}, S.~{Aggarwal} et~al., \emph{{Measurement of
  High-energy Cosmic-Ray Proton Spectrum from the ISS-CREAM Experiment}},
  \href{https://doi.org/10.3847/1538-4357/ac9d2c}{\emph{\apj} {\bfseries 940}
  (2022) 107}.

\bibitem{2019ApJ...871...96A}
A.U.~{Abeysekara}, R.~{Alfaro}, C.~{Alvarez} et~al., \emph{{All-sky Measurement
  of the Anisotropy of Cosmic Rays at 10 TeV and Mapping of the Local
  Interstellar Magnetic Field}},
  \href{https://doi.org/10.3847/1538-4357/aaf5cc}{\emph{\apj} {\bfseries 871}
  (2019) 96}.

\bibitem{2012ApJ...752...68V}
A.E.~{Vladimirov}, G.~{J{\'o}hannesson}, I.V.~{Moskalenko} et~al.,
  \emph{{Testing the Origin of High-energy Cosmic Rays}},
  \href{https://doi.org/10.1088/0004-637X/752/1/68}{\emph{\apj} {\bfseries 752}
  (2012) 68}.

\bibitem{2012PhRvL.109f1101B}
P.~{Blasi}, E.~{Amato}, P.D.~{Serpico}, \emph{{Spectral Breaks as a Signature
  of Cosmic Ray Induced Turbulence in the Galaxy}},
  \href{https://doi.org/10.1103/PhysRevLett.109.061101}{\emph{\prl} {\bfseries
  109} (2012) 061101}.

\bibitem{2023FrPhy..1844301M}
P.-X.~{Ma}, Z.-H.~{Xu}, Q.~{Yuan} et~al., \emph{{Interpretations of the cosmic
  ray secondary-to-primary ratios measured by DAMPE}},
  \href{https://doi.org/10.1007/s11467-023-1257-7}{\emph{Frontiers of Physics}
  {\bfseries 18} (2023) 44301}.

\bibitem{2019JCAP...10..010L}
W.~{Liu}, Y.-Q.~{Guo}, Q.~{Yuan}, \emph{{Indication of nearby source signatures
  of cosmic rays from energy spectra and anisotropies}},
  \href{https://doi.org/10.1088/1475-7516/2019/10/010}{\emph{\jcap} {\bfseries
  2019} (2019) 010}.

\bibitem{2020ApJ...903...69F}
K.~{Fang}, X.-J.~{Bi}, P.-F.~{Yin}, \emph{{DAMPE Proton Spectrum Indicates a
  Slow-diffusion Zone in the nearby ISM}},
  \href{https://doi.org/10.3847/1538-4357/abb8d7}{\emph{\apj} {\bfseries 903}
  (2020) 69}.

\bibitem{2021PhRvD.104j3013F}
O.~{Fornieri}, D.~{Gaggero}, D.~{Guberman} et~al., \emph{{Diffusive origin for
  the cosmic-ray spectral hardening reveals signatures of a nearby source in
  the leptons and protons data}},
  \href{https://doi.org/10.1103/PhysRevD.104.103013}{\emph{\prd} {\bfseries
  104} (2021) 103013}.

\bibitem{2021FrPhy..1624501Y}
Q.~{Yuan}, B.-Q.~{Qiao}, Y.-Q.~{Guo} et~al., \emph{{Nearby source
  interpretation of differences among light and medium composition spectra in
  cosmic rays}},
  \href{https://doi.org/10.1007/s11467-020-0990-4}{\emph{Frontiers of Physics}
  {\bfseries 16} (2021) 24501}.

\bibitem{2021PhRvD.104l3001Z}
M.-J.~{Zhao}, K.~{Fang}, X.-J.~{Bi}, \emph{{Constraints on the spatially
  dependent cosmic-ray propagation model from Bayesian analysis}},
  \href{https://doi.org/10.1103/PhysRevD.104.123001}{\emph{\prd} {\bfseries
  104} (2021) 123001}.

\bibitem{2022ApJ...926...41Z}
B.~{Zhao}, W.~{Liu}, Q.~{Yuan} et~al., \emph{{Geminga SNR: Possible Candidate
  of the Local Cosmic-Ray Factory}},
  \href{https://doi.org/10.3847/1538-4357/ac4416}{\emph{\apj} {\bfseries 926}
  (2022) 41}.

\bibitem{2022ApJ...930...82L}
Q.~{Luo}, B.-q.~{Qiao}, W.~{Liu} et~al., \emph{{Statistical Study of the
  Optimal Local Sources for Cosmic Ray Nuclei and Electrons}},
  \href{https://doi.org/10.3847/1538-4357/ac6267}{\emph{\apj} {\bfseries 930}
  (2022) 82}.

\bibitem{2023Galax..11...43L}
Q.~{Luo}, J.~{Feng}, P.-H.T.~{Tam}, \emph{{Explaining the Hardening Structures
  of Helium Spectrum and Boron to Carbon Ratio through Different Propagation
  Models}}, \href{https://doi.org/10.3390/galaxies11020043}{\emph{Galaxies}
  {\bfseries 11} (2023) 43}.

\bibitem{2022arXiv221205641Q}
B.-Q.~{Qiao}, Y.-Q.~{Guo}, {Wei-Liu} et~al., \emph{{Nearby SNR: a possible
  common origin to multi-messenger anomalies in spectra, ratios and anisotropy
  of cosmic rays}},
  \href{https://doi.org/10.48550/arXiv.2212.05641}{\emph{arXiv e-prints} (2022)
  arXiv:2212.05641}.

\bibitem{2023ApJ...942...13Q}
B.-Q.~{Qiao}, Q.~{Luo}, Q.~{Yuan} et~al., \emph{{Understanding the Phase
  Reversals of Galactic Cosmic-Ray Anisotropies}},
  \href{https://doi.org/10.3847/1538-4357/aca7fc}{\emph{\apj} {\bfseries 942}
  (2023) 13}.

\bibitem{2022MNRAS.511.6218Z}
Y.~{Zhang}, S.~{Liu}, H.~{Zeng}, \emph{{A three-component model for cosmic ray
  spectrum and dipole anisotropy}},
  \href{https://doi.org/10.1093/mnras/stac470}{\emph{\mnras} {\bfseries 511}
  (2022) 6218}.

\bibitem{2023JCAP...02..007Z}
P.-P.~{Zhang}, X.-Y.~{He}, W.~{Liu} et~al., \emph{{Evidence of fresh cosmic ray
  in galactic plane based on DAMPE measurement of B/C and B/O ratios}},
  \href{https://doi.org/10.1088/1475-7516/2023/02/007}{\emph{\jcap} {\bfseries
  2023} (2023) 007}.

\bibitem{2023ApJ...952..100N}
L.~{Nie}, Y.~{Liu}, Z.~{Jiang}, \emph{{Implications of a Possible Spectral
  Structure of Cosmic-Ray Protons Unveiled by the DAMPE}},
  \href{https://doi.org/10.3847/1538-4357/acda29}{\emph{\apj} {\bfseries 952}
  (2023) 100}.

\bibitem{2015PhRvD..92h1301T}
N.~{Tomassetti}, \emph{{Cosmic-ray protons, nuclei, electrons, and
  antiparticles under a two-halo scenario of diffusive propagation}},
  \href{https://doi.org/10.1103/PhysRevD.92.081301}{\emph{\prd} {\bfseries 92}
  (2015) 081301}.

\bibitem{2000A&A...362..937A}
F.A.~{Aharonian}, A.M.~{Atoyan}, \emph{{Broad-band diffuse gamma ray emission
  of the galactic disk}},
  \href{https://doi.org/10.48550/arXiv.astro-ph/0009009}{\emph{\aap} {\bfseries
  362} (2000) 937}.

\bibitem{2006PhRvD..74c4018K}
S.R.~{Kelner}, F.A.~{Aharonian}, V.V.~{Bugayov}, \emph{{Energy spectra of gamma
  rays, electrons, and neutrinos produced at proton-proton interactions in the
  very high energy regime}},
  \href{https://doi.org/10.1103/PhysRevD.74.034018}{\emph{\prd} {\bfseries 74}
  (2006) 034018}.

\bibitem{2021ApJ...911..151M}
M.A.~{Malkov}, I.V.~{Moskalenko}, \emph{{The TeV Cosmic-Ray Bump: A Message
  from the Epsilon Indi or Epsilon Eridani Star?}},
  \href{https://doi.org/10.3847/1538-4357/abe855}{\emph{\apj} {\bfseries 911}
  (2021) 151}.

\bibitem{2022ApJ...933...78M}
M.A.~{Malkov}, I.V.~{Moskalenko}, \emph{{On the Origin of Observed Cosmic-Ray
  Spectrum Below 100 TV}},
  \href{https://doi.org/10.3847/1538-4357/ac7049}{\emph{\apj} {\bfseries 933}
  (2022) 78}.

\bibitem{Malkov:2023fQ}
M.~Malkov, \emph{{On Why the 10-TeV Cosmic Ray Bump Originates in the Local
  Interstellar Medium}},  in \emph{\ICRCthirtyeight {\textemdash}
  PoS(ICRC2023)}, vol.~444, 2023, 143.

\bibitem{1979ICRC....1..488N}
J.~{Nishimura}, M.~{Fujii}, T.~{Taira}, \emph{{Electron Spectrum of the High
  Energy Side}},  in \emph{\ICRCsixteen}, vol.~1, 1979, p.488.

\bibitem{1995A&A...294L..41A}
F.A.~{Aharonian}, A.M.~{Atoyan}, H.J.~{Voelk}, \emph{{High energy electrons and
  positrons in cosmic rays as an indicator of the existence of a nearby cosmic
  tevatron}}, {\emph{\aap} {\bfseries 294} (1995) L41}.

\bibitem{2004ApJ...601..340K}
T.~{Kobayashi}, Y.~{Komori}, K.~{Yoshida} et~al., \emph{{The Most Likely
  Sources of High-Energy Cosmic-Ray Electrons in Supernova Remnants}},
  \href{https://doi.org/10.1086/380431}{\emph{\apj} {\bfseries 601} (2004)
  340}.

\bibitem{2009PhRvL.102r1101A}
A.A.~{Abdo}, M.~{Ackermann}, M.~{Ajello} et~al., \emph{{Measurement of the
  Cosmic Ray e$^{+}$+e$^{-}$ Spectrum from 20 GeV to 1 TeV with the Fermi Large
  Area Telescope}},
  \href{https://doi.org/10.1103/PhysRevLett.102.181101}{\emph{\prl} {\bfseries
  102} (2009) 181101}.

\bibitem{2010PhRvD..82i2004A}
M.~{Ackermann}, M.~{Ajello}, W.B.~{Atwood} et~al., \emph{{Fermi LAT
  observations of cosmic-ray electrons from 7 GeV to 1 TeV}},
  \href{https://doi.org/10.1103/PhysRevD.82.092004}{\emph{\prd} {\bfseries 82}
  (2010) 092004}.

\bibitem{Adriani:2023SR}
O.~Adriani, Y.~Akaike, K.~Asano et~al., \emph{{The cosmic-ray electron and
  positron spectrum measured with CALET on the International Space Station}},
  in \emph{\ICRCthirtyeight {\textemdash} PoS(ICRC2023)}, vol.~444, 2023, 071.

\bibitem{2008PhRvL.101z1104A}
F.~{Aharonian}, A.G.~{Akhperjanian}, U.~{Barres de Almeida} et~al.,
  \emph{{Energy Spectrum of Cosmic-Ray Electrons at TeV Energies}},
  \href{https://doi.org/10.1103/PhysRevLett.101.261104}{\emph{\prl} {\bfseries
  101} (2008) 261104}.

\bibitem{2009A&A...508..561A}
F.~{Aharonian}, A.G.~{Akhperjanian}, G.~{Anton} et~al., \emph{{Probing the ATIC
  peak in the cosmic-ray electron spectrum with H.E.S.S.}},
  \href{https://doi.org/10.1051/0004-6361/200913323}{\emph{\aap} {\bfseries
  508} (2009) 561}.

\bibitem{2018PhRvD..98f2004A}
A.~{Archer}, W.~{Benbow}, R.~{Bird} et~al., \emph{{Measurement of cosmic-ray
  electrons at TeV energies by VERITAS}},
  \href{https://doi.org/10.1103/PhysRevD.98.062004}{\emph{\prd} {\bfseries 98}
  (2018) 062004}.

\bibitem{2017RNCim.40...473A}
O.~{Adriani}, G.C.~{Barbarino}, G.A.~{Bazilevskaya} et~al., \emph{{Ten Years of
  PAMELA in Space}}, {\emph{Riv.\ Nuovo Cimento} {\bfseries 40} (2017) 473}.

\bibitem{2014PhRvL.113v1102A}
M.~{Aguilar}, D.~{Aisa}, B.~{Alpat} et~al., \emph{{Precision Measurement of the
  (e$^{+}$+e$^{-}$) Flux in Primary Cosmic Rays from 0.5 GeV to 1 TeV with the
  Alpha Magnetic Spectrometer on the International Space Station}},
  \href{https://doi.org/10.1103/PhysRevLett.113.221102}{\emph{\prl} {\bfseries
  113} (2014) 221102}.

\bibitem{2019PhRvL.122j1101A}
M.~{Aguilar}, L.~{Ali Cavasonza}, B.~{Alpat} et~al., \emph{{Towards
  Understanding the Origin of Cosmic-Ray Electrons}},
  \href{https://doi.org/10.1103/PhysRevLett.122.101101}{\emph{\prl} {\bfseries
  122} (2019) 101101}.

\bibitem{2018PhRvL.120z1102A}
O.~{Adriani}, Y.~{Akaike}, K.~{Asano} et~al., \emph{{Extended Measurement of
  the Cosmic-Ray Electron and Positron Spectrum from 11 GeV to 4.8 TeV with the
  Calorimetric Electron Telescope on the International Space Station}},
  \href{https://doi.org/10.1103/PhysRevLett.120.261102}{\emph{\prl} {\bfseries
  120} (2018) 261102}.

\bibitem{2017Natur.552...63D}
G.~{Ambrosi}, Q.~{An}, R.~{Asfandiyarov} et~al., \emph{{Direct detection of a
  break in the teraelectronvolt cosmic-ray spectrum of electrons and
  positrons}}, \href{https://doi.org/10.1038/nature24475}{\emph{\nat}
  {\bfseries 552} (2017) 63}.

\bibitem{2017PhRvD..95h2007A}
S.~{Abdollahi}, M.~{Ackermann}, M.~{Ajello} et~al., \emph{{Cosmic-ray
  electron-positron spectrum from 7 GeV to 2 TeV with the Fermi Large Area
  Telescope}}, \href{https://doi.org/10.1103/PhysRevD.95.082007}{\emph{\prd}
  {\bfseries 95} (2017) 082007}.

\bibitem{2017Sci...358..911A}
A.U.~{Abeysekara}, A.~{Albert}, R.~{Alfaro} et~al., \emph{{Extended gamma-ray
  sources around pulsars constrain the origin of the positron flux at Earth}},
  \href{https://doi.org/10.1126/science.aan4880}{\emph{Science} {\bfseries 358}
  (2017) 911}.

\bibitem{2017JCAP...01..006M}
S.~{Manconi}, M.~{Di Mauro}, F.~{Donato}, \emph{{Dipole anisotropy in cosmic
  electrons and positrons: inspection on local sources}},
  \href{https://doi.org/10.1088/1475-7516/2017/01/006}{\emph{\jcap} {\bfseries
  2017} (2017) 006}.

\bibitem{Motz:2023AX}
H.~Motz, O.~Adriani, Y.~Akaike et~al., \emph{{Interpretation of the CALET
  Electron+Positron Spectrum by Astrophysical Sources}},  in
  \emph{\ICRCthirtyeight {\textemdash} PoS(ICRC2023)}, vol.~444, 2023, 067.

\bibitem{Kounine:2023AV}
A.~Kounine, \emph{{Understanding the Origin of Cosmic-Ray Electrons}},  in
  \emph{\ICRCthirtyeight {\textemdash} PoS(ICRC2023)}, vol.~444, 2023, 065.

\bibitem{1996ApJ...457L.103G}
R.L.~{Golden}, S.J.~{Stochaj}, S.A.~{Stephens} et~al., \emph{{Measurement of
  the Positron to Electron Ratio in Cosmic Rays above 5 GeV}},
  \href{https://doi.org/10.1086/309896}{\emph{\apjl} {\bfseries 457} (1996)
  L103}.

\bibitem{2004PhRvL..93x1102B}
J.J.~{Beatty}, A.~{Bhattacharyya}, C.~{Bower} et~al., \emph{{New Measurement of
  the Cosmic-Ray Positron Fraction from 5 to 15 GeV}},
  \href{https://doi.org/10.1103/PhysRevLett.93.241102}{\emph{\prl} {\bfseries
  93} (2004) 241102}.

\bibitem{2009Natur.458..607A}
O.~{Adriani}, G.C.~{Barbarino}, G.A.~{Bazilevskaya} et~al., \emph{{An anomalous
  positron abundance in cosmic rays with energies 1.5-100 GeV}},
  \href{https://doi.org/10.1038/nature07942}{\emph{\nat} {\bfseries 458} (2009)
  607}.

\bibitem{1982ApJ...254..391P}
R.J.~{Protheroe}, \emph{{On the nature of the cosmic ray positron spectrum}},
  \href{https://doi.org/10.1086/159743}{\emph{\apj} {\bfseries 254} (1982)
  391}.

\bibitem{1998ApJ...493..694M}
I.V.~{Moskalenko}, A.W.~{Strong}, \emph{{Production and Propagation of
  Cosmic-Ray Positrons and Electrons}},
  \href{https://doi.org/10.1086/305152}{\emph{\apj} {\bfseries 493} (1998)
  694}.

\bibitem{2012PhRvL.108a1103A}
M.~{Ackermann}, M.~{Ajello}, A.~{Allafort} et~al., \emph{{Measurement of
  Separate Cosmic-Ray Electron and Positron Spectra with the Fermi Large Area
  Telescope}},
  \href{https://doi.org/10.1103/PhysRevLett.108.011103}{\emph{\prl} {\bfseries
  108} (2012) 011103}.

\bibitem{2014PhRvL.113l1101A}
L.~{Accardo}, M.~{Aguilar}, D.~{Aisa} et~al., \emph{{High Statistics
  Measurement of the Positron Fraction in Primary Cosmic Rays of 0.5-500 GeV
  with the Alpha Magnetic Spectrometer on the International Space Station}},
  \href{https://doi.org/10.1103/PhysRevLett.113.121101}{\emph{\prl} {\bfseries
  113} (2014) 121101}.

\bibitem{2019PhRvL.122d1102A}
M.~{Aguilar}, L.~{Ali Cavasonza}, G.~{Ambrosi} et~al., \emph{{Towards
  Understanding the Origin of Cosmic-Ray Positrons}},
  \href{https://doi.org/10.1103/PhysRevLett.122.041102}{\emph{\prl} {\bfseries
  122} (2019) 041102}.

\bibitem{2016PhRvL.117i1103A}
M.~{Aguilar}, L.~{Ali Cavasonza}, B.~{Alpat} et~al., \emph{{Antiproton Flux,
  Antiproton-to-Proton Flux Ratio, and Properties of Elementary Particle Fluxes
  in Primary Cosmic Rays Measured with the Alpha Magnetic Spectrometer on the
  International Space Station}},
  \href{https://doi.org/10.1103/PhysRevLett.117.091103}{\emph{\prl} {\bfseries
  117} (2016) 091103}.

\bibitem{Weng:20231Q}
Z.~Weng, \emph{{Antiproton Flux and Properties of Elementary Particle Fluxes in
  Primary Cosmic Rays Measured with the Alpha Magnetic Spectrometer on the
  ISS}},  in \emph{\ICRCthirtyeight {\textemdash} PoS(ICRC2023)}, vol.~444,
  2023, 1403.

\bibitem{2017PhRvD..95f3009L}
P.~{Lipari}, \emph{{Interpretation of the cosmic ray positron and antiproton
  fluxes}}, \href{https://doi.org/10.1103/PhysRevD.95.063009}{\emph{\prd}
  {\bfseries 95} (2017) 063009}.

\bibitem{2019PhRvD..99d3005L}
P.~{Lipari}, \emph{{Spectral shapes of the fluxes of electrons and positrons
  and the average residence time of cosmic rays in the Galaxy}},
  \href{https://doi.org/10.1103/PhysRevD.99.043005}{\emph{\prd} {\bfseries 99}
  (2019) 043005}.

\bibitem{2023ApJ...950..120C}
I.~{Cholis}, I.~{Krommydas}, \emph{{Possible Counterpart Signal of the Fermi
  Bubbles at the Cosmic-Ray Positrons}},
  \href{https://doi.org/10.3847/1538-4357/accb55}{\emph{\apj} {\bfseries 950}
  (2023) 120}.

\bibitem{2010ApJ...724.1044S}
M.~{Su}, T.R.~{Slatyer}, D.P.~{Finkbeiner}, \emph{{Giant Gamma-ray Bubbles from
  Fermi-LAT: Active Galactic Nucleus Activity or Bipolar Galactic Wind?}},
  \href{https://doi.org/10.1088/0004-637X/724/2/1044}{\emph{\apj} {\bfseries
  724} (2010) 1044}.

\bibitem{2014ApJ...793...64A}
M.~{Ackermann}, A.~{Albert}, W.B.~{Atwood} et~al., \emph{{The Spectrum and
  Morphology of the Fermi Bubbles}},
  \href{https://doi.org/10.1088/0004-637X/793/1/64}{\emph{\apj} {\bfseries 793}
  (2014) 64}.

\bibitem{Calore:2023AV}
F.~Calore, \emph{{Dark matter searches: status and prospects}},  in
  \emph{\ICRCthirtyeight {\textemdash} PoS(ICRC2023)}, vol.~444, 2023, 025.

\bibitem{2009PhRvL.103e1104B}
P.~{Blasi}, \emph{{Origin of the Positron Excess in Cosmic Rays}},
  \href{https://doi.org/10.1103/PhysRevLett.103.051104}{\emph{\prl} {\bfseries
  103} (2009) 051104}.

\bibitem{2003A&A...410..189B}
E.G.~{Berezhko}, L.T.~{Ksenofontov}, V.S.~{Ptuskin} et~al., \emph{{Cosmic ray
  production in supernova remnants including reacceleration: The secondary to
  primary ratio}},
  \href{https://doi.org/10.1051/0004-6361:20031274}{\emph{\aap} {\bfseries 410}
  (2003) 189}.

\bibitem{2009PhRvL.103h1103B}
P.~{Blasi}, P.D.~{Serpico}, \emph{{High-Energy Antiprotons from Old Supernova
  Remnants}}, \href{https://doi.org/10.1103/PhysRevLett.103.081103}{\emph{\prl}
  {\bfseries 103} (2009) 081103}.

\bibitem{2009PhRvL.103h1104M}
P.~{Mertsch}, S.~{Sarkar}, \emph{{Testing Astrophysical Models for the PAMELA
  Positron Excess with Cosmic Ray Nuclei}},
  \href{https://doi.org/10.1103/PhysRevLett.103.081104}{\emph{\prl} {\bfseries
  103} (2009) 081104}.

\bibitem{2013PhRvD..87d7301K}
M.~{Kachelrie{\ss}}, S.~{Ostapchenko}, \emph{{B/C ratio and the PAMELA positron
  excess}}, \href{https://doi.org/10.1103/PhysRevD.87.047301}{\emph{\prd}
  {\bfseries 87} (2013) 047301}.

\bibitem{2014PhRvD..89d3013C}
I.~{Cholis}, D.~{Hooper}, \emph{{Constraining the origin of the rising cosmic
  ray positron fraction with the boron-to-carbon ratio}},
  \href{https://doi.org/10.1103/PhysRevD.89.043013}{\emph{\prd} {\bfseries 89}
  (2014) 043013}.

\bibitem{2016PhRvD..94f3006M}
M.A.~{Malkov}, P.H.~{Diamond}, R.Z.~{Sagdeev}, \emph{{Positive charge
  prevalence in cosmic rays: Room for dark matter in the positron spectrum}},
  \href{https://doi.org/10.1103/PhysRevD.94.063006}{\emph{\prd} {\bfseries 94}
  (2016) 063006}.

\bibitem{1981IAUS...94..175A}
J.~{Arons}, \emph{{Particle acceleration by pulsars}},  in \emph{Origin of
  Cosmic Rays}, G.~{Setti}, G.~{Spada}, A.W.~{Wolfendale}, eds., vol.~94, 1981,
  p.175.

\bibitem{1987ICRC....2...92H}
A.K.~{Harding}, R.~{Ramaty}, \emph{{The Pulsar Contribution to Galactic Cosmic
  Ray Positrons}},  in \emph{\ICRCtwenty}, vol.~2, 1987, p.92.

\bibitem{1989ApJ...342..807B}
A.~{Boulares}, \emph{{The Nature of the Cosmic-Ray Electron Spectrum, and
  Supernova Remnant Contributions}},
  \href{https://doi.org/10.1086/167637}{\emph{\apj} {\bfseries 342} (1989)
  807}.

\bibitem{2013ApJ...772...18L}
T.~{Linden}, S.~{Profumo}, \emph{{Probing the Pulsar Origin of the Anomalous
  Positron Fraction with AMS-02 and Atmospheric Cherenkov Telescopes}},
  \href{https://doi.org/10.1088/0004-637X/772/1/18}{\emph{\apj} {\bfseries 772}
  (2013) 18}.

\bibitem{2017PhRvD..96j3013H}
D.~{Hooper}, I.~{Cholis}, T.~{Linden} et~al., \emph{{HAWC observations strongly
  favor pulsar interpretations of the cosmic-ray positron excess}},
  \href{https://doi.org/10.1103/PhysRevD.96.103013}{\emph{\prd} {\bfseries 96}
  (2017) 103013}.

\bibitem{2017SSRv..207..235B}
A.M.~{Bykov}, E.~{Amato}, A.E.~{Petrov} et~al., \emph{{Pulsar Wind Nebulae with
  Bow Shocks: Non-thermal Radiation and Cosmic Ray Leptons}},
  \href{https://doi.org/10.1007/s11214-017-0371-7}{\emph{\ssr} {\bfseries 207}
  (2017) 235}.

\bibitem{2019ApJ...876L...8B}
A.M.~{Bykov}, A.E.~{Petrov}, A.M.~{Krassilchtchikov} et~al., \emph{{GeV-TeV
  Cosmic-Ray Leptons in the Solar System from the Bow Shock Wind Nebula of the
  Nearest Millisecond Pulsar J0437-4715}},
  \href{https://doi.org/10.3847/2041-8213/ab1922}{\emph{\apjl} {\bfseries 876}
  (2019) L8}.

\bibitem{1979PhRvL..43.1196G}
R.L.~{Golden}, S.~{Horan}, B.G.~{Mauger} et~al., \emph{{Evidence for the
  Existence of Cosmic-Ray Antiprotons}},
  \href{https://doi.org/10.1103/PhysRevLett.43.1196}{\emph{\prl} {\bfseries 43}
  (1979) 1196}.

\bibitem{1979ICRC....1..330B}
E.A.~{Bogomolov}, N.D.~{Lubyanaya}, V.A.~{Romanov} et~al., \emph{{A
  Stratospheric Magnetic Spectrometer Investigation of the Singly Charged
  Component Spectra and Composition of the Primary and Secondary Cosmic
  Radiation}},  in \emph{\ICRCsixteen}, vol.~1, 1979, p.330.

\bibitem{2013NuPhS.243...92Y}
A.~{Yamamoto}, J.W.~{Mitchell}, \emph{{Search for Primary Antiparticles and
  Cosmological Antimatter with BESS}},
  \href{https://doi.org/10.1016/j.nuclphysbps.2013.09.027}{\emph{\npbps}
  {\bfseries 243} (2013) 92}.

\bibitem{1996ApJ...467L..33H}
M.~{Hof}, W.~{Menn}, C.~{Pfeifer} et~al., \emph{{Measurement of Cosmic-Ray
  Antiprotons from 3.7 to 19 GeV}},
  \href{https://doi.org/10.1086/310185}{\emph{\apjl} {\bfseries 467} (1996)
  L33}.

\bibitem{2001PhRvL..87A1101B}
A.S.~{Beach}, J.J.~{Beatty}, A.~{Bhattacharyya} et~al., \emph{{Measurement of
  the Cosmic-Ray Antiproton-to-Proton Abundance Ratio between 4 and 50 GeV}},
  \href{https://doi.org/10.1103/PhysRevLett.87.271101}{\emph{\prl} {\bfseries
  87} (2001) 271101}.

\bibitem{2001ApJ...561..787B}
M.~{Boezio}, V.~{Bonvicini}, P.~{Schiavon} et~al., \emph{{The Cosmic-Ray
  Antiproton Flux between 3 and 49 GeV}},
  \href{https://doi.org/10.1086/323366}{\emph{\apj} {\bfseries 561} (2001)
  787}.

\bibitem{2010PhRvL.105l1101A}
O.~{Adriani}, G.C.~{Barbarino}, G.A.~{Bazilevskaya} et~al., \emph{{PAMELA
  Results on the Cosmic-Ray Antiproton Flux from 60 MeV to 180 GeV in Kinetic
  Energy}}, \href{https://doi.org/10.1103/PhysRevLett.105.121101}{\emph{\prl}
  {\bfseries 105} (2010) 121101}.

\bibitem{1998ApJ...499..250S}
M.~{Simon}, A.~{Molnar}, S.~{Roesler}, \emph{{A New Calculation of the
  Interstellar Secondary Cosmic-Ray Antiprotons}},
  \href{https://doi.org/10.1086/305606}{\emph{\apj} {\bfseries 499} (1998)
  250}.

\bibitem{2017ApJ...840..115B}
M.J.~{Boschini}, S.~{Della Torre}, M.~{Gervasi} et~al., \emph{{Solution of
  Heliospheric Propagation: Unveiling the Local Interstellar Spectra of
  Cosmic-ray Species}},
  \href{https://doi.org/10.3847/1538-4357/aa6e4f}{\emph{\apj} {\bfseries 840}
  (2017) 115}.

\bibitem{2017PhRvL.118s1101C}
M.-Y.~{Cui}, Q.~{Yuan}, Y.-L.S.~{Tsai} et~al., \emph{{Possible Dark Matter
  Annihilation Signal in the AMS-02 Antiproton Data}},
  \href{https://doi.org/10.1103/PhysRevLett.118.191101}{\emph{\prl} {\bfseries
  118} (2017) 191101}.

\bibitem{2017PhRvL.118s1102C}
A.~{Cuoco}, M.~{Kr{\"a}mer}, M.~{Korsmeier}, \emph{{Novel Dark Matter
  Constraints from Antiprotons in Light of AMS-02}},
  \href{https://doi.org/10.1103/PhysRevLett.118.191102}{\emph{\prl} {\bfseries
  118} (2017) 191102}.

\bibitem{2006ApJ...642..902P}
V.S.~{Ptuskin}, I.V.~{Moskalenko}, F.C.~{Jones} et~al., \emph{{Dissipation of
  Magnetohydrodynamic Waves on Energetic Particles: Impact on Interstellar
  Turbulence and Cosmic-Ray Transport}},
  \href{https://doi.org/10.1086/501117}{\emph{\apj} {\bfseries 642} (2006)
  902}.

\bibitem{2020PhRvR...2d3017H}
J.~{Heisig}, M.~{Korsmeier}, M.W.~{Winkler}, \emph{{Dark matter or correlated
  errors: Systematics of the AMS-02 antiproton excess}},
  \href{https://doi.org/10.1103/PhysRevResearch.2.043017}{\emph{\prr}
  {\bfseries 2} (2020) 043017}.

\bibitem{2020PhRvD.102j3007E}
N.E.~{Engelbrecht}, V.~{Di Felice}, \emph{{Uncertainties implicit to the use of
  the force-field solutions to the Parker transport equation in analyses of
  observed cosmic ray antiproton intensities}},
  \href{https://doi.org/10.1103/PhysRevD.102.103007}{\emph{\prd} {\bfseries
  102} (2020) 103007}.

\bibitem{2021ApJ...908..167E}
N.E.~{Engelbrecht}, K.D.~{Moloto}, \emph{{An Ab Initio Approach to Antiproton
  Modulation in the Inner Heliosphere}},
  \href{https://doi.org/10.3847/1538-4357/abd3a5}{\emph{\apj} {\bfseries 908}
  (2021) 167}.

\bibitem{2023arXiv230400760L}
X.-J.~{Lv}, X.-J.~{Bi}, K.~{Fang} et~al., \emph{{Reanalysis of the Systematic
  Uncertainties in Cosmic-Ray Antiproton Flux}},
  \href{https://doi.org/10.48550/arXiv.2304.00760}{\emph{arXiv e-prints} (2023)
  arXiv:2304.00760}.

\bibitem{2019ApJ...880...95K}
C.M.~{Karwin}, S.~{Murgia}, S.~{Campbell} et~al., \emph{{Fermi-LAT Observations
  of {\ensuremath{\gamma}}-Ray Emission toward the Outer Halo of M31}},
  \href{https://doi.org/10.3847/1538-4357/ab2880}{\emph{\apj} {\bfseries 880}
  (2019) 95}.

\bibitem{2021PhRvD.103b3027K}
C.M.~{Karwin}, S.~{Murgia}, I.V.~{Moskalenko} et~al., \emph{{Dark matter
  interpretation of the Fermi-LAT observations toward the outer halo of M31}},
  \href{https://doi.org/10.1103/PhysRevD.103.023027}{\emph{\prd} {\bfseries
  103} (2021) 023027}.

\bibitem{2015ApJ...803...54K}
M.~{Kachelriess}, I.V.~{Moskalenko}, S.S.~{Ostapchenko}, \emph{{New Calculation
  of Antiproton Production by Cosmic Ray Protons and Nuclei}},
  \href{https://doi.org/10.1088/0004-637X/803/2/54}{\emph{\apj} {\bfseries 803}
  (2015) 54}.

\bibitem{2023CoPhC.28708698K}
M.~{Kachelrie{\ss}}, S.~{Ostapchenko}, J.~{Tjemsland}, \emph{{AAfrag 2.01:
  interpolation routines for Monte Carlo results on secondary production
  including light antinuclei in hadronic interactions}},
  \href{https://doi.org/10.1016/j.cpc.2023.108698}{\emph{\cpc} {\bfseries 287}
  (2023) 108698}.

\bibitem{2017JCAP...02..048W}
M.W.~{Winkler}, \emph{{Cosmic ray antiprotons at high energies}},
  \href{https://doi.org/10.1088/1475-7516/2017/02/048}{\emph{\jcap} {\bfseries
  2017} (2017) 048}.

\bibitem{2017PhRvD..96d3007D}
F.~{Donato}, M.~{Korsmeier}, M.~{Di Mauro}, \emph{{Prescriptions on antiproton
  cross section data for precise theoretical antiproton flux predictions}},
  \href{https://doi.org/10.1103/PhysRevD.96.043007}{\emph{\prd} {\bfseries 96}
  (2017) 043007}.

\bibitem{2018PhRvD..97j3019K}
M.~{Korsmeier}, F.~{Donato}, M.~{Di Mauro}, \emph{{Production cross sections of
  cosmic antiprotons in the light of new data from the NA61 and LHCb
  experiments}}, \href{https://doi.org/10.1103/PhysRevD.97.103019}{\emph{\prd}
  {\bfseries 97} (2018) 103019}.

\bibitem{2018PhRvL.121v2001A}
R.~{Aaij}, C.~{Abell{\'a}n Beteta}, B.~{Adeva} et~al., \emph{{Measurement of
  Antiproton Production in p-He Collisions at $\sqrt{s_{NN}}$=110 GeV}},
  \href{https://doi.org/10.1103/PhysRevLett.121.222001}{\emph{\prl} {\bfseries
  121} (2018) 222001}.

\bibitem{2020PhRvR...2b3022B}
M.~{Boudaud}, Y.~{G{\'e}nolini}, L.~{Derome} et~al., \emph{{AMS-02 antiprotons'
  consistency with a secondary astrophysical origin}},
  \href{https://doi.org/10.1103/PhysRevResearch.2.023022}{\emph{Physical Review
  Research} {\bfseries 2} (2020) 023022}.

\bibitem{2019PhRvD..99j3026C}
I.~{Cholis}, T.~{Linden}, D.~{Hooper}, \emph{{A robust excess in the cosmic-ray
  antiproton spectrum: Implications for annihilating dark matter}},
  \href{https://doi.org/10.1103/PhysRevD.99.103026}{\emph{\prd} {\bfseries 99}
  (2019) 103026}.

\bibitem{2021MPLA...3630003H}
J.~{Heisig}, \emph{{Cosmic-ray antiprotons in the AMS-02 era: A sensitive probe
  of dark matter}},
  \href{https://doi.org/10.1142/S0217732321300032}{\emph{\mpla} {\bfseries 36}
  (2021) 2130003}.

\bibitem{2022ScPP...12..163C}
F.~{Calore}, M.~{Cirelli}, L.~{Derome} et~al., \emph{{AMS-02 antiprotons and
  dark matter: Trimmed hints and robust bounds}},
  \href{https://doi.org/10.21468/SciPostPhys.12.5.163}{\emph{SciPost Physics}
  {\bfseries 12} (2022) 163}.

\bibitem{2022PhRvL.129w1101Z}
C.-R.~{Zhu}, M.-Y.~{Cui}, Z.-Q.~{Xia} et~al., \emph{{Explaining the GeV
  Antiproton Excess, GeV {\ensuremath{\gamma}}-Ray Excess, and W-Boson Mass
  Anomaly in an Inert Two Higgs Doublet Model}},
  \href{https://doi.org/10.1103/PhysRevLett.129.231101}{\emph{\prl} {\bfseries
  129} (2022) 231101}.

\bibitem{2023arXiv230706798G}
Y.~{G{\'e}nolini}, D.~{Maurin}, I.V.~{Moskalenko} et~al., \emph{{Current status
  and desired accuracy of the isotopic production cross-sections relevant to
  astrophysics of cosmic rays II. Fluorine to Silicon (and updated LiBeB)}},
  \href{https://doi.org/10.48550/arXiv.2307.06798}{\emph{arXiv e-prints} (2023)
  arXiv:2307.06798}.

\end{thebibliography}\endgroup

\end{document}